\documentclass[twocolumn,usenames,dvipsnames,twocolappendix]{aastex63}

\usepackage{amsmath}

\shorttitle{Galactic Disk Formation in the Early Universe}
\shortauthors{Semenov et al.}
\graphicspath{{./}{figures/}}

\usepackage{etoolbox}
\makeatletter
\patchcmd{\NAT@citex}
  {\@citea\NAT@hyper@{\NAT@nmfmt{\NAT@nm}\NAT@date}}
  {\@citea\NAT@nmfmt{\NAT@nm}\NAT@hyper@{\NAT@date}}
  {}
  {}
\patchcmd{\NAT@citex}
  {\@citea\NAT@hyper@{%
     \NAT@nmfmt{\NAT@nm}%
     \hyper@natlinkbreak{\NAT@aysep\NAT@spacechar}{\@citeb\@extra@b@citeb}%
     \NAT@date}}
  {\@citea\NAT@nmfmt{\NAT@nm}%
   \NAT@aysep\NAT@spacechar%
   \NAT@hyper@{\NAT@date}}
  {}
  {}
\patchcmd{\NAT@citex}
  {\@citea\NAT@hyper@{%
     \NAT@nmfmt{\NAT@nm}%
     \hyper@natlinkbreak{\NAT@spacechar\NAT@@open\if*#1*\else#1\NAT@spacechar\fi}%
       {\@citeb\@extra@b@citeb}%
     \NAT@date}}
  {\@citea\NAT@nmfmt{\NAT@nm}%
   \NAT@spacechar\NAT@@open\if*#1*\else#1\NAT@spacechar\fi%
   \NAT@hyper@{\NAT@date}}
  {}
  {}
\makeatother

\def\Rvir{R_{\rm vir}}

\def\cs{c_{\rm s}}

\def\Mstar{M_{\rm \star}}

\def\Zii{Z_{\rm II}}
\def\Zia{Z_{\rm Ia}}

\def\jzjc{j_z/j_{\rm c}}

\def\vt{v_\tau}
\def\vcirc{v_{\rm circ}}

\def\rhoSFR{\dot{\rho}_{\star}}
\def\SFR{\dot{M}_{\star}}
\def\sfr{\dot{m}_{\star}}
\def\tff{t_{\rm ff}}

\def\epsff{\epsilon_{\rm ff}}
\def\avir{\alpha_{\rm vir}}
\def\stot{\sigma_{\rm tot}}
\def\st{\sigma_{\rm turb}}
\def\eturb{e_{\rm turb}}

\def\pc{{\rm \;pc}}
\def\kpc{{\rm \;kpc}}

\def\kms{{\rm \;km\;s^{-1}}}
\def\Msunyr{{\rm \;M_\odot\;yr^{-1}}}
\def\Msun{{\rm \;M_\odot}}
\def\Zsun{{\rm \;Z_\odot}}
\def\Myr{{\rm \;Myr}}
\def\Gyr{{\rm \;Gyr}}
\def\cc{{\rm \;cm^{-3}}}
\def\K{{\rm \;K}}

\newcommand{\newtext}{}

\begin{document}

\title{How Early Could the Milky Way's Disk Form?}

\author[0000-0002-6648-7136]{Vadim A. Semenov}
\altaffiliation{\href{mailto:vadim.semenov@cfa.harvard.edu}{vadim.semenov@cfa.harvard.edu}}
\affiliation{Center for Astrophysics $|$ Harvard \& Smithsonian, 60 Garden St, Cambridge, MA 02138, USA}

\author[0000-0002-1590-8551]{Charlie Conroy}
\affiliation{Center for Astrophysics $|$ Harvard \& Smithsonian, 60 Garden St, Cambridge, MA 02138, USA}

\author[0000-0002-2838-9033]{Aaron Smith}
\affiliation{Department of Physics, The University of Texas at Dallas, Richardson, TX 75080, USA}

\author[0000-0001-8778-7587]{Ewald Puchwein}
\affiliation{Leibniz-Institut f\"ur Astrophysik Potsdam, An der Sternwarte 16, 14482 Potsdam, Germany}

\author[0000-0001-6950-1629]{Lars Hernquist}
\affiliation{Center for Astrophysics $|$ Harvard \& Smithsonian, 60 Garden St, Cambridge, MA 02138, USA}

\begin{abstract}
We investigate early, $z > 3$, galaxy formation in a cosmological zoom-in simulation of a close, early-forming Milky Way (MW) analog extracted from TNG50 simulation and resimulated with detailed modeling of cold interstellar medium (ISM) formation, coupled with on-the-fly UV radiative transfer, turbulence-regulated star formation, and stellar feedback. In our enhanced-physics simulation, the galaxy develops a bistable ISM structure (warm, with $T \sim 10^4\K$, and cold, with $T < 100\K$) and exhibits significantly more efficient, early, and bursty star formation than in TNG. Notably, the stellar disk of this MW progenitor forms extremely early, around $z\sim6\text{--}7$, and exhibits chemo-kinematic properties consistent with the low-metallicity population of the MW stars. The disk forms rapidly, on a timescale of $\sim$0.2 Gyr which is significantly shorter than the timescale implied by the observable chemo-kinematic signatures of disk spinup, $\sim$0.7 Gyr, due to the scatter in the age--metallicity relation. The rotational support of the gas disk and the location of the galaxy on the main sequence are consistent with early disk galaxies observed by \emph{JWST} and ALMA at $z\sim4\text{--}7$, suggesting that some of these galaxies could be progenitors of MW-like systems. Remarkably, the variation of the global star formation rate (SFR) before disk formation is similar to the observed SFR scatter at these early times. Our findings underscore the critical role of modeling a turbulent cold ISM and turbulence-regulated star formation and feedback in driving early SFR variability, while at the same time enabling early disk formation, without destroying it with overly efficient stellar feedback.
\end{abstract}


\section{Introduction}

Within its first two years, the \emph{James Webb Space Telescope} (\emph{JWST}) has fundamentally transformed our paradigm of galaxy formation. The first results revealed a surprising abundance of bright galaxies in the early Universe \citep[e.g.,][]{finkelstein22,naidu22,boylan-kolchin23,labbe23}. Although these findings appear to be in tension with the theoretical models of galaxy evolution, some existing cosmological simulations and analytical models produced abundances of bright galaxies consistent with the \emph{JWST} results as long as the variability of star formation rates (SFRs) at early times was accounted for \citep[e.g.,][]{keller23,mccaffrey23,sun23-fire,shen23,kravtsov-belokurov24}. Whether this high SFR variability is physical or is too large, e.g., due to overly efficient feedback, remains a subject of active debate. 

Several alternative explanations for this high abundance of early galaxies have been proposed in the literature, that among others include stellar feedback being inefficient at early times \citep[e.g.,][]{dekel23,qin23,boylan-kolchin24}, and variations in the initial mass function (IMF) shape \citep[e.g.,][]{shen23,yung24} and dust content \citep[e.g.,][]{ferrara23}. Regardless of the specific details, the point stands that, in the early Universe, galaxies form and evolve qualitatively differently from the local Universe.

Apart from the high abundance of early bright galaxies, results from \emph{JWST} \citep[e.g.,][]{ferreira22a,ferreira22b,jacobs22,naidu22,nelson22,robertson23} and earlier from ALMA \citep[e.g.,][]{smit18,neeleman20,pensabene20,rizzo20,rizzo21,fraternali21,lelli21,tsukui-iguchi21,herreracamus22,posses23,romanoliveira23} showed that dynamically cold galactic disks can form significantly earlier than previously thought. Galaxies observed with \emph{JWST} exhibit predominantly disk morphologies out to $z\sim8$ \citep[][]{ferreira22b}. If these galaxies are indeed disks, this will imply that disks form very early and persist throughout cosmic time \citep[see, however,][]{gibson24,pandya24}.
Closer to home, the local Galactic archeological data also suggest that the disk of our own Milky Way formed surprisingly early, within the first $\sim$2 Gyr of evolution \citep{bk22,conroy22,rix22,xiang-rix22,nepal24}, which appears to be in tension with the results of state-of-the-art cosmological simulations of MW-mass galaxies, which typically form disks several Gyr later \citep{bk22,mccluskey23}. 

These findings make the first 2 Gyr of the evolution of the Universe, i.e., $z > 3$, interesting in the context of early disk assembly and the transition from early chaotic and bursty galaxy formation to steady star-forming disk galaxies typical of the present-day Universe. Modeling galaxy formation at these early times is particularly challenging because simulations often rely on extrapolation of subgrid models calibrated against local observations to the extreme regimes typical for the early Universe: high densities and SFRs, high levels of turbulence, low metallicity, and, early on, the absence of a well-defined disk.

Accurately modeling the processes of star formation and feedback is crucial for disk formation, as has been shown in a large body of literature over the past two decades \citep[e.g.,][]{governato04,governato07,okamoto05,robertson06,scannapieco08,zavala08,brook12,bird13,bird21,fire,fire2,ubler14,agertz15,agertz16,genel15,christensen16,auriga,el-badry18,garrison-kimmel18,pillepich19,vintergatan1,vintergatan2,vintergatan3,gurvich22,hafen22,khoperskov22a,khoperskov22c,rodriguez-gomez22,vintergatan4,yu22,semenov23a}. Stellar feedback plays multiple different roles in disk formation. First, it expels low-angular momentum gas from galaxies, preventing its accumulation in the center and buildup of bulge-dominated systems \citep[e.g.,][]{agertz16}. Second, by keeping the disk mass fraction low, feedback stabilizes disks against global gravitational instabilities that could also drive material into the center and cause disk fragmentation \citep{efstathiou82,mo-mao-white98}. Third, by powering the matter cycle through the circumgalactic medium, recycled outflows can bring in high-angular momentum gas from the halo outskirts, promoting disk growth \citep[e.g.,][]{ubler14,christensen16,defelippis17,semenov23b}. Finally, overly strong feedback can substantially thicken or even destroy the disk \citep[e.g.,][]{agertz16}.

The last effect may explain why current state-of-the-art zoom-in simulations of MW-like galaxies struggle to reproduce early disk formation \citep{bk22}. While efficient feedback can induce significant variation in the SFR at early times in the galaxy assembly process, alleviating the tension with the observed abundances of early UV-bright galaxies, the same effect can delay or hinder the formation of galactic disks at these high redshifts \citep[e.g.,][]{dekel20,gurvich22}. For example, while FIRE-2 simulations are consistent with the \emph{JWST} luminosity functions up to $z \sim 12$ owing to the efficient and bursty nature of implemented stellar feedback \citep{sun23-fire}, the galactic disks in MW-mass progenitors form significantly later than suggested by Galactic archeological data \citep[at $z \sim 1\text{--}2$;][]{mccluskey23}.

On the other hand, the early formation of the MW could be attributed to the unusually rapid assembly of its dark matter halo \citep{dillamore24,semenov23a}. As was shown by \citet{semenov23a} based on the TNG50\footnote{TNG50 is the highest resolution run of the IllustrisTNG cosmological simulation suite \citep{nelson19,pillepich19}. In the rest of the text, we will use ``TNG50'' and ``TNG'' to refer to the simulation itself and the underlying galaxy formation model summarized in Section~\ref{sec:methods:tng}, respectively.} cosmological simulation, although on average MW-mass disks indeed form relatively late (consistent with the results of zoom-in simulations), the variation in disk formation times is substantial, so that $\sim10\%$ of such galaxies form disks as early as the MW data suggests. However, in TNG50, the modeling of star formation and feedback relies on the effective equation of state approach, which tends to produce significantly smoother star formation histories compared to high-resolution zoom-in simulations that explicitly model the formation of a cold ISM. The reduced burstiness is one of the reasons why such simulations underpredict the number of bright galaxies at $z > 10$ \citep{kannan23}. Thus, the interplay between feedback modeling and assembly history in galactic disk formation and evolution remains an open question.

To explore this issue, we investigate the formation and evolution of a close MW analog at $z > 3$ in cosmological simulations with detailed modeling of a cold ISM, star formation, and feedback. Specifically, we extract one of the early-forming MW-like disk galaxies from the TNG50 simulation that exhibits a remarkably similar chemo-kinematic structure of the stellar disk to observations at $z=0$ suggesting a mass assembly history similar to that of the MW \citep{chandra23}. We resimulate this galaxy using a high-resolution cosmological zoom-in setup with explicit modeling of cooling and heating coupled with a self-consistently evolving UV radiation field and a physically motivated model for turbulence-regulated star formation and stellar feedback developed and thoroughly tested in \cite{semenov16,semenov18,semenov21-tf}.  This model has successfully reproduced the observed near-linear correlation between molecular gas and SFR on $>$kpc scales without imposing such a relation locally \citep{semenov17,semenov19}, as well as their spatial decoupling on subkiloparsec scales \citep{semenov21-tf}. The latter statistic is particularly challenging to capture in galaxy simulations, making it a sensitive probe of the star formation and feedback modeling \citep{semenov18,fujimoto19,chevance23}. Recently, \citet{polzin23,polzin24} also explored the model in simulations of low-metallicity dwarf disk galaxies, extending down to $\sim 0.01\Zsun$. The fact that such an agreement is achieved without tuning and instead with forward-modeling ISM properties like unresolved turbulent energy and local star formation efficiencies makes this model particularly appealing for exploring the effects of star formation and feedback on early galaxy evolution, where subgrid model calibration is difficult.

Our paper is structured as follows. In Section~\ref{sec:methods}, we describe the choice of the initial conditions and outline the numerical galaxy formation models used in our study. Section~\ref{sec:results} presents the key findings by contrasting the results of our new simulation with the original TNG model in terms of the global gas distribution, star formation histories, and early galactic disk formation and evolution. In particular, in Sections~\ref{sec:results:mw-data} and \ref{sec:results:jwst-alma}, we show that the new simulation is broadly consistent with both the local observations of low-metallicity stars in the MW and high-redshift disk galaxies observed by \emph{JWST} and ALMA. We discuss our findings in Section~\ref{sec:discussion} and summarize our conclusions in Section~\ref{sec:summary}. Appendices~\ref{app:resolution} and \ref{app:FeH} provide additional details on the resolution of our simulations, \newtext{including convergence tests,} and the calculation of metal abundances, respectively.

\section{Methods}
\label{sec:methods}

\begin{deluxetable*}{lcc}
\tablecaption{Key differences between the models
\label{tab:models}}
\tablewidth{0pt}
\tablehead{
\colhead{\bf Property} &
\colhead{\bf ART} &
\colhead{\bf TNG}
}
\startdata
Hydrodynamics framework & Adaptive mesh refinement & Moving mesh \\
Baryon resolution & $\sim 10^{4} \Msun$; $\Delta x_{\rm med} \approx 25\pc$\tablenotemark{\scriptsize a} & $\sim 8.5 \times 10^{4} \Msun$; $\Delta x_{\rm med} \approx 100\pc$ \\
ISM thermodynamics & \parbox{6cm}{\centering Explicit $Z$- and UV-dependent \\ cooling down to few K} & \parbox{6cm}{\centering Effective EoS with \\ $T \sim 10^4\text{--}10^5\K$ at $n > 0.1\cc$\tablenotemark{\scriptsize b}} \\
Heating sources & \parbox{6cm}{\centering On-the-fly UV RT (background + stars) \\ + dissipation of subgrid turbulence} & Uniform UV background + AGN \\
Nonthermal pressure & Subgrid turbulent pressure & --- \\
Star formation prescription & Locally variable $\epsff(\avir)$ & KS-like relation calibrated at $z=0$ \\
Stellar feedback & Adaptive energy and momentum injection & Effective EoS + decoupled winds \\
AGN feedback & --- & Thermal + kinetic modes \\
Molecular chemistry & Nonequilibrium network & --- \\
\enddata
\tablenotetext{\scriptsize a}{The cited mass resolution corresponds to the gas mass refinement criterion (see the text and Figure~\ref{fig:resolution} in Appendix~\ref{app:resolution:comparison}). The cell size at a given refinement level is fixed in comoving units and the cited value is provided at $z=3$, where most of the comparison in this paper is done. $\Delta x_{\rm med}$ is the median cell size in the galaxy ISM, defined as all gas cells at densities greater than $n>0.3\cc$.}
\tablenotetext{\scriptsize b}{See Figure~\ref{fig:n-T}.}
\end{deluxetable*}

In this section, we describe how the initial conditions for our simulations of an early-forming MW analog were selected (Section~\ref{sec:methods:ics}), and outline the key features of the two versions of our runs: the new simulation with detailed cold ISM physics run with the ART code, which we henceforth label ``ART'' (Section~\ref{sec:methods:art}) and the original TNG model (Section~\ref{sec:methods:tng}). The main differences between the two models are summarized in Table~\ref{tab:models}.

\subsection{Galaxy Model: Early-Forming MW Analog}
\label{sec:methods:ics} 

To both investigate early galaxy evolution and draw a plausible picture of the early formation of the MW's disk, we select a prototypical MW analog from the TNG50 cosmological simulation studied by \citet[][subfind halo ID {\tt 519311} at $z = 0$]{chandra23}. This is one of the early-forming MW-mass galaxies identified in \citet{semenov23b,semenov23a}, which shows the formation of a stellar disk at lower than average metallicities, [Fe/H] $< -1$, consistent with the observational measurements by \citet{bk22}, and additionally \newtext{exhibits} structures in the [Fe/H] vs. $\jzjc$ plane (such as the thick and thin disks and the transitions between them) which are remarkably similar to the MW data \citep[][]{chandra23}.
The halo mass of this galaxy at $z=0$ and $z=3$---when we perform most of our analysis below---is $M_{\rm 200c} \approx 1.4\times10^{12}\Msun$ and $3.0\times10^{11}\Msun$, respectively.

To generate the initial conditions (ICs) for zoom-in resimulations of this galaxy from the original TNG50 ICs (a representative $\sim 50^3$ comoving Mpc$^3$ region of the universe, assuming Planck XIII \citeyear{planck-xiii} cosmology), the resolution of the dark matter particles was degraded outside of the Lagrangian volume of its halo (the numerical details of this procedure will be described in \citealt{puchwein24}). The dark matter particle resolution within the Lagrangian region of our target halo was kept the same as that in TNG50, $4.5 \times 10^5\Msun$. At $z \approx 3$, when we perform most of our analyses, no low-resolution dark matter particles reside within the virial radius in the TNG run, while in the ART run, they contribute $<0.3\%$ of the total dark matter mass. This difference originates from the back-reaction of cold and dense gas on the dark matter structure formation.

One of the differences of our target halo from other early-forming MW analogs in TNG50 is that it experiences two significant mergers at $z \sim 3$ and $z \sim 0.7$ \citep[see Figure~8 in][]{chandra23}. While with the TNG model, the disk survives through these mergers, this is not the case with the detailed ISM modeling in ART, which we find by resimulating the targeted halo at a lower resolution past these mergers. For this reason, we perform our analysis at $z \approx 3$, prior to such a destructive merger occurring in ART. We discuss the potential effects of the ISM modeling on mergers and disk survivability in Section~\ref{sec:discussion} and leave a detailed study for future work.

\subsection{ART Overview}
\label{sec:methods:art}

To investigate the effects of realistic cold ISM modeling on the formation of galactic disks in the early universe, we resimulate the above MW analog from TNG50 using the adaptive mesh refinement $N$-body and hydrodynamics code ART \citep{kravtsov99,kravtsov02,rudd08,gnedin11} with on-the-fly modeling of UV radiation transfer \citep{gnedin14} and using the ISM, star formation, and feedback model that was developed and thoroughly tested in \citet{semenov16,semenov17,semenov18,semenov19,semenov21-tf}. In MW-mass galaxies, this model produces a realistic Kennicutt--Schmidt (KS) relation for total and molecular gas, in terms of both normalization and slope without imposing such a relation locally \citep{semenov17,semenov19} and allows us to make meaningful predictions for the local properties of star-forming gas at resolution scales thanks to a physically motivated model for turbulence-regulated star formation efficiency \citep{semenov18,semenov21-tf}. In particular, as shown in \citet{semenov21-tf}, this model is able, without tuning, to reproduce the observed spatial decorrelation of young stars and dense gas on subkiloparsec scales---the statistic that was shown to be a sensitive probe of the star formation--feedback cycle modeling \citep{semenov18,fujimoto19,chevance23}. This model was also tested in low-metallicity (down to 0.01 solar) regimes by \citet{polzin23,polzin24}. All this makes the model particularly appealing for application in simulations of early galaxy formation where more common models calibrated against $z=0$ observations necessarily rely on extrapolation.
The key improvements in our new simulation compared to the original TNG model are summarized in Table~\ref{tab:models}. Here, we briefly outline these differences and for a more detailed description of the model refer to \citet[][]{semenov21-tf}.

We adopt a higher baryon resolution in our ART simulation than in the original TNG50 run, motivated by the typical resolution of the idealized galaxy simulations where our model has been extensively tested (see the references above). Specifically, the AMR grid is adaptively refined, when the total gas mass in a cell exceeds $\sim 5.7 \times 10^4 \Msun$, thereby keeping the median cell gas mass on different levels roughly constant, $\sim 10^4 \Msun$, which is a factor of $\sim 8$ smaller than in TNG, albeit with a significantly larger scatter (see Figure~\ref{fig:resolution} in Appendix~\ref{app:resolution:comparison}). The mesh is refined in this fashion down to the minimal cell size of $\sim 100$ comoving pc, or $\sim 25$ pc at $z=3$. 
The star particle mass resolution in ART ($10^4 \Msun$) is also higher than in TNG ($\sim 8.5 \times 10^4 \Msun$), allowing for better temporal and sampling resolution in the stellar properties discussed throughout this paper.
\newtext{The gas resolution of our simulation is chosen to be close to that in the previous idealized simulations, where our star formation and feedback model has been extensively tested \citep[$\Delta x_{\rm min} \sim 10\text{--}40\pc$; e.g.,][]{semenov17,semenov18,semenov21-tf}. We present the comparison of resolutions in our ART and TNG runs in Appendix~\ref{app:resolution:comparison} and demonstrate that our results are robust for the chosen resolution in Appendix~\ref{app:resolution:convergence}.}

The gravitational potential is calculated by using a Fast Fourier Transform at the lowest grid level and relaxation method on all higher refinement levels, with the effective resolution for gravity corresponding to $\sim$2--4 cells \citep[see][]{kravtsov97,gnedin16,mansfield21}. 
The gravitational potential used to accelerate dark matter particles is computed at 3 levels above the highest-resolution level, corresponding to the cell size of $\sim 800$ comoving pc, while the potential for stars and gas is reconstructed at the highest-resolution level, corresponding to $\sim 100$ comoving pc. We have checked that increasing and decreasing the gravity resolution for dark matter particles have only a small impact on the dark matter halo assembly, especially compared to the effects from the formation of dense and cold gas.

Apart from differences in hydrodynamics framework and resolution, the most important difference between our new simulation and TNG is the modeling of ISM physics, which includes realistic cooling and heating dependent on metallicity and a local incident UV radiation field \citep{gnedin12}, on-the-fly transfer of UV radiation \newtext{produced by both the local sources and the \citet{haardt12} cosmological background} \citep[using the Optically Thin Variable Eddington Tensor, OTVET, approximation;][]{gnedin01,gnedin14}, and a dynamic model for unresolved turbulence (using the ``shear-improved'' Large-Eddy Simulation model of \citealt{schmidt14} as described in \citealt{semenov16}).
As we show below, with these changes, the ISM gas develops a complex multiphase structure, while the unresolved turbulent energy provides nonthermal pressure support in each cell and is used to modulate the local star formation.

We parameterize the local star formation rate in each cell via the star formation efficiency per freefall time, $\tff = \sqrt{3\pi/32G\rho}$:
\begin{equation}
\label{eq:rhosfr}
\rhoSFR = \epsff \frac{\rho}{\tff} \, .
\end{equation}
Galaxy simulations often treat $\epsff$ as a tunable parameter and introduce star formation thresholds calibrated against available observations. In our simulation, we instead allow $\epsff$ to vary continuously with the local properties of unresolved turbulence, enabling us to make meaningful predictions about the distribution and evolution of $\epsff$ at high redshifts. Specifically, in our simulation, $\epsff$ depends on the (subgrid) virial parameter, $\avir$, following the fit to magnetohydrodynamic simulations of turbulent star-forming regions by \citet{padoan12}:
\begin{equation}
\label{eq:epsff-P12}
\epsff = 0.9 \exp{(-\sqrt{\avir/0.53})} \, ,
\end{equation}
with the choice of the prefactor explained in \citet{semenov16} and the virial parameter defined as for a uniform sphere with the radius equal to half of the cell size, $R = \Delta/2$ \citep{bertoldi92}:
\begin{equation}
\label{eq:avir}
    \avir \equiv \frac{5 \stot^2 R}{3GM} \approx 6 \frac{ (\stot/5\kms)^2 }{ (n/100\cc) (\Delta/25 \pc)^2} \, ,
\end{equation}
where $\stot = \sqrt{\st^2+\cs^2}$ accounts for both the unresolved turbulent velocity dispersion, $\st = \sqrt{2\,\eturb/\rho}$, and thermal support, expressed via the sound speed, $\cs$. The small-scale turbulent velocity dispersion is calculated from the explicitly followed subgrid turbulent energy, $\eturb$, which is sourced by the local cascade of kinetic energy from resolved scales and dissipates into heat on the local cell-crossing time \citep[see][for details]{semenov16}.

Young stars affect the ISM by sourcing UV radiation that is self-consistently propagated by the RT solver and via injection of thermal energy and radial momentum following our fiducial model from \citet{semenov21-tf}. The amount of energy and radial momentum injected per supernova (SN) depends on local gas density, metallicity, and cell size, using results from simulations of SN remnants expanding into a nonuniform ISM by \citet{martizzi15}. In our fiducial model, we additionally boost the radial momentum by a factor of 5 to account for the effects of SN clustering \citep[e.g.,][]{gentry17,gentry19} and cosmic-ray pressure \citep{diesing18}, both of which can increase the injected momentum by a factor of a few. 
\newtext{In addition to SNe, early feedback from stellar winds, unresolved \ion{H}{2} regions, and radiation pressure injects a similar amount of momentum per unit time as SNe while acting before the first SN occurs \citep[e.g.,][]{agertz13}. To approximate this behavior, we compute the energy and momentum as described above and initiate the injection immediately upon the formation of the stellar particle, without the typical delay of a few Myr corresponding to the lifetime of the most massive SNII progenitors.}
This approximation neglects the metallicity-dependence of early feedback \citep[e.g.,][]{dekel23}, however, we find that, in our model, early feedback has only a weak effect on the global SFR when the local $\epsff$ is not too high, so that only a small fraction of star-forming region is converted into stars before the first SN explodes \citep{semenov18,semenov21-tf}.
The total number of SNe for a given star particle is computed using the \citet{chabrier03} IMF.

Unlike TNG, the ART run does not include active galactic nuclei (AGN) feedback. As we explicitly demonstrate below, at early times when we perform our analyses, our target galaxy is not massive enough for AGN feedback to produce a significant effect, and therefore, the absence of AGN feedback in ART cannot explain the differences between the runs that we report below.

During their evolution, stellar particles return a fraction of their mass following the prescription of \cite{leitner11}.
In ART, we separately follow the total mass fractions of metals injected by SNe type II and Ia, $\Zii$ and $\Zia$, which we use to reconstruct [Fe/H] assuming the TNG yield tables \citep{pillepich18} as described in Appendix~\ref{app:FeH}.

We also explicitly follow the formation and destruction of molecular gas using a nonequilibrium six species chemical network from \citet{gnedin11} coupled with the locally variable UV radiation field. The molecular gas is modeled independently from star formation, which allows us to make predictions for the relation between molecular and star-forming gas in the early Universe, which we do in a companion paper \citep{semenov24b}.

\subsection{TNG Overview}
\label{sec:methods:tng} 

To facilitate a direct comparison with the original TNG model, we rerun the same zoom-in ICs with the model parameters as the TNG50 simulation but with a higher output cadence matching that of the ART simulation. 

The simulation was carried out using the moving-mesh $N$-body and magnetohydrodynamic code Arepo \citep{arepo}, where the unstructured mesh based on Voronoi tessellation of the domain continuously moves and deforms following the flow in quasi-Lagrangian fashion, with the mass of gas cells maintained within a factor of $\sim$2 from the target resolution, $8.5 \times 10^4\Msun$. The resulting mass-weighted median cell size in the ISM gas with $n > 0.3\cc$ at $z=3$ is $\sim 100\pc$ (see Appendix~\ref{app:resolution:comparison} and Figure~\ref{fig:resolution}). The gravitational softening length for dark matter, stellar, and wind particles is 575 comoving pc at $z>1$ and 288 physical pc at $z<1$, while for gas, the softening length varies with the cell size as $\epsilon_{\rm gas} = 2.5\;r_{\rm cell}$, where $r_{\rm cell}$ is the radius of a sphere with the same volume as the Voronoi cell.

In TNG, the gas with densities $n < 0.1\cc$ is subject to radiative cooling (primordial, Compton, and metal-line) and heating from a spatially uniform \citet{faucher-giguere09} UV background, which is turned on at $z=6$ \newtext{and is locally attenuated according to the \citet{rahmati13} gas self-shielding model}, and by local AGN sources using the approximate method described in \citet{vogelsberger13}. At densities $n > 0.1\cc$, the gas temperature and pressure are set by the \citet{sh03} effective equation of state, with modifications described in \citet{vogelsberger13} and \citet{nelson19}.

The local star formation rate, which is used to stochastically convert gas cells into star particles, is calibrated against the observed Kennicutt--Schmidt relation from \citet{kennicutt98}. The corresponding local star formation efficiency per freefall time, $\epsff = \sfr/(m/\tff)$, varies smoothly and very weakly as a function of density between $\epsff \approx 3.5\%$ and 4\%---due to the variation of the cold gas fraction in the \citet{sh03} model---and therefore can be considered fixed.

Galaxy scale stellar feedback is modeled by stochastically ejecting locally prescribed ISM gas mass in the form of decoupled wind particles, which are recoupled with the background gas after they leave the galaxy \citep{pillepich18,nelson19}. Unlike ART, the TNG model also includes AGN feedback, which is modeled in two modes, based on the local Eddington-limited Bondi gas accretion rate: continuous injection of thermal energy and stochastic kinetic kicks at high and low accretion rates, respectively \citep{weinberger17}.

The TNG simulation explicitly tracks 9 chemical elements, including H and Fe, which can be used to compare with the observed stellar abundances, [Fe/H]. In ART, on the other hand, we do not follow individual species and instead follow the total mass fractions of metals injected by SNe type II and Ia, $\Zii$ and $\Zia$---quantities that are also available in TNG. To make the comparison between TNG and ART consistent, we derive [Fe/H] from $\Zii$ and $\Zia$ using TNG yield tables \citep{pillepich18} as described in Appendix~\ref{app:FeH}.

\section{Results}
\label{sec:results}

In this section, we present the results of our zoom-in resimulation of an MW-analog from TNG50 using the detailed ISM, star formation, and feedback modeling in the ART code and compare it with the results from the TNG model. As described in Section~\ref{sec:methods:ics}, we present our analysis at $z \approx 3$, which is approximately the time when the final disk forms in this galaxy in the original TNG50
simulation and before the destructive merger that destroys the disk in the ART simulation.

\subsection{Formation of Dense and Cold ISM}
\label{sec:results:dense-gas}

\begin{figure*}
\centering
\includegraphics[width=0.5\textwidth]{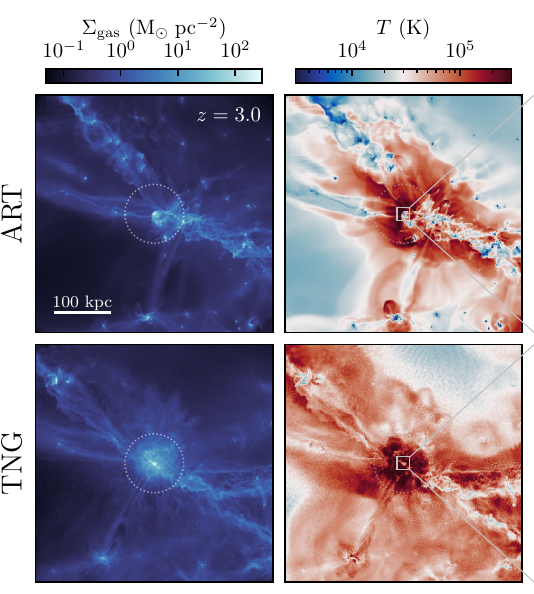}%
\includegraphics[width=0.5\textwidth]{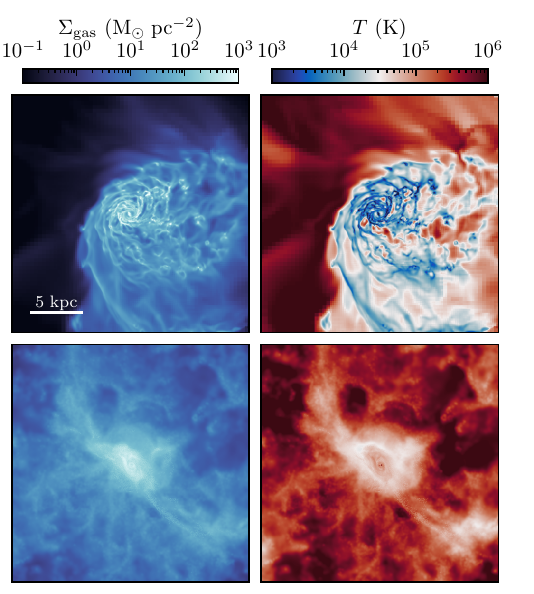}
\caption{\label{fig:maps} Maps of gas surface density and mass-weighted temperature along the line of sight in our ART (top) and TNG simulations (bottom) at $z=3$. The left set of panels show the gas structure on large scales, while the right set of panels zooms in to the galactic disk. The dotted circle indicates the virial radius defined using the redshift-dependent spherical overdensity from \citet{bryan-norman98}. With the detailed modeling of cold ISM, the ART simulation exhibits significantly more dense cold structures on small scales and develops a significantly more prominent thin gaseous disk than in TNG.}
\end{figure*}

Figure~\ref{fig:maps} compares the spatial distributions of the gas density and temperature at $z=3$, both on large scales (the left set of panels) and zooming in at the galaxy scale (the right set of panels). The large-scale features, such as the locations of dense filaments and the presence of hot ($T > 10^5\K$) gas, are qualitatively similar in both runs. In detail, however, the distributions differ significantly.

The most striking difference is the presence of dense and cold structures in the ART simulation, owing to explicit modeling of cooling down to low temperatures. On large scales, the ART simulation shows significantly more dense and cold ``blobs,'' early star-forming dwarf galaxies. On small scales, the main progenitor also appears qualitatively different: in ART, it develops a thin extended disk with a prominent spiral structure, while in TNG, the galaxy is significantly smaller, puffier, and featureless. 

\begin{figure}
\centering
\includegraphics[width=\columnwidth]{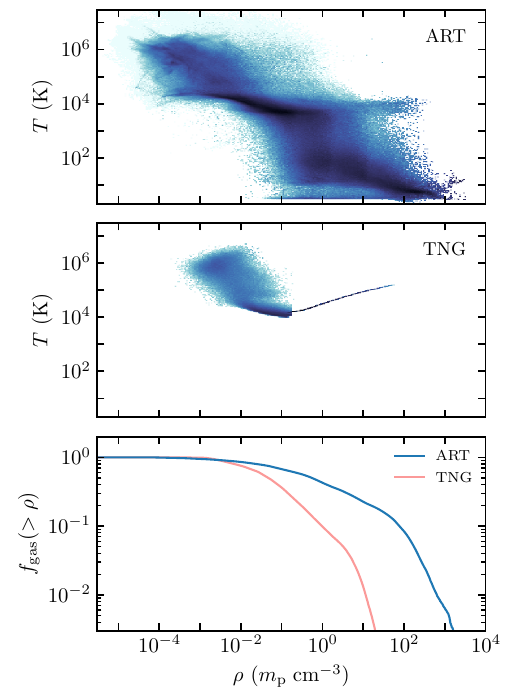}
\caption{\label{fig:n-T} Mass-weighted joint distribution of gas density and temperature in our ART (top panel) and TNG (middle panel) simulations at $z=3$, and the comparison of cumulative PDFs (bottom panel). Only the gas within $30\kpc$ from the galaxy center, excluding the central 0.5 kpc, is shown. The detailed modeling of cooling and heating in ART leads to the formation of a two-phase medium at densities of $n \gtrsim 0.1\cc$ with temperatures $T \sim 10^4\K$ (WNM) and $T \lesssim 100\K$ (CNM). In TNG, gas thermodynamics at these densities is replaced with an effective equation of state with significantly higher $T$ (and therefore pressure), than in ART. Such a stiff equation of state prevents the formation of dense gas in TNG compared to ART as shown in the bottom panel.}
\end{figure}

\begin{figure}
\centering
\includegraphics[width=\columnwidth]{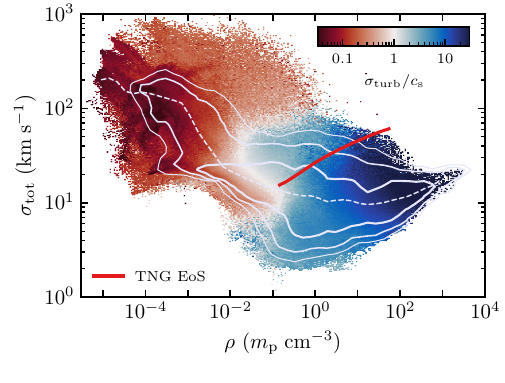}
\caption{\label{fig:n-sigma} Joint distribution of gas density and total velocity dispersion, $\stot = \sqrt{\st^2+\cs^2}$, including the contributions of thermal pressure and subgrid turbulence, in the ART simulation. The color shows the mass-weighted average subgrid Mach number in each pixel, $\st/\cs$, while blue contours enclose 68\%, 95\%, and 99\% of gas mass, and the dashed line shows a running median. In the dense, supersonic ISM (the blue portion of the PDF) the median $\stot$ in ART is near constant, implying a roughly isothermal effective equation of state (EoS), albeit with significant scatter. For comparison, the red solid line shows the effective EoS imposed in TNG, which is significantly stiffer, leading to a factor of $\sim$30--40 higher effective pressure, $P \propto \rho \stot^2$, in dense gas with $n\sim100\cc$.}
\end{figure}

These differences stem from the treatments of gas thermodynamics at high densities as quantified in Figures~\ref{fig:n-T} and \ref{fig:n-sigma}. Figure~\ref{fig:n-T} compares the density and temperature distribution of gas, while Figure~\ref{fig:n-sigma} shows the relation between the density and total velocity dispersion, $\stot = \sqrt{\st^2+\cs^2}$, including the contributions of thermal pressure and subgrid turbulence in ART. The latter figure is analogous to the $n$--$T$ plot except that it quantifies the total pressure support in the dense, supersonic gas where subgrid turbulence dominates the pressure support, as indicated by the color. To compare the local pressure support of dense gas between ART and TNG, $\stot$ in ART can be directly compared with the sound speed corresponding to the effective equation of state imposed in TNG (shown with the solid red line). In the plots, we show the gas within 30 kpc from the galaxy center and exclude the central 0.5 kpc to avoid the high-density and high-$\st$ artifact in ART caused by strong shear in the very center \citep[see Figure~8 in ][]{semenov22}.

As Figure~\ref{fig:n-T} shows, owing to explicit cooling in ART, the gas at $n>0.1\cc$, develops a two-phase medium with $T \sim 10^4\K$ (warm neutral medium, WNM; with a contribution of resolved \ion{H}{2} regions at high densities) and $T \lesssim 100\K$ (cold neutral medium; CNM). In TNG, on the other hand, the thermodynamics of gas at such densities is replaced with the \citet{sh03} effective equation of state. In this model, the effective temperature and pressure are derived as that of a mixture of hot ($T \sim 10^6\K$) and ``cold'' ($T \sim 10^3\K$, representing a mixture of WNM and CNM) phases, and the resulting temperature is significantly higher than that in the ART simulation. 

Indeed, as Figure~\ref{fig:n-sigma} shows, in gas with $n\sim100\cc$, the effective sound speed in TNG is a factor of $\sim$5--6 larger than in ART, implying that the effective pressure, $P \propto \rho \stot^2$, is $\sim$30--40 times higher. We stress again that the pressure in ART is significantly lower despite it including the contribution from small-scale turbulence, which dominates over the thermal pressure at these densities, i.e., the dense ISM is supersonic (see the color in the figure). The significantly higher pressure in TNG prevents the formation of dense structures. For comparison, in TNG, the main progenitor hardly has any gas at $n>10\cc$, while in ART, more than 10\% of the gas within $\Rvir$ is at $n>100\cc$ (see the bottom panel in Figure~\ref{fig:n-T}).

\subsection{Early, Efficient, and Bursty Star Formation}

\begin{figure}
\centering
\includegraphics[width=\columnwidth]{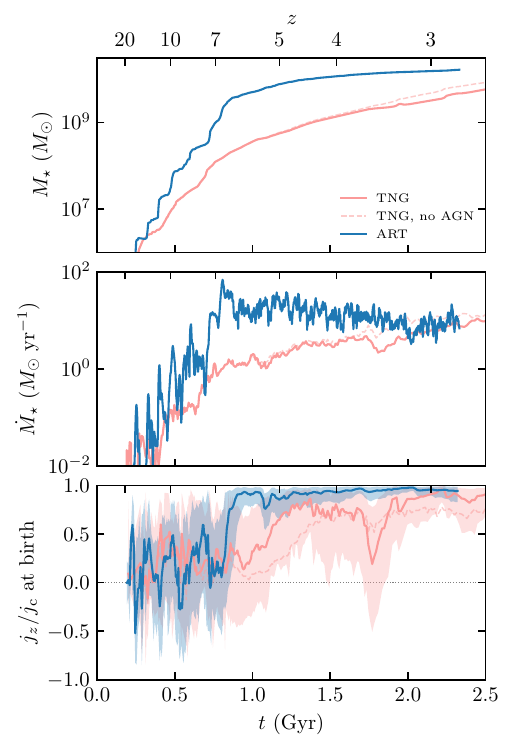}
\caption{\label{fig:sfh-jzjc} Total stellar mass ($\Mstar$; top panel), global \newtext{in-situ} SFR ($\SFR$; middle panel), and the orbital circularity of young, $<100\Myr$, stars ($\jzjc$; bottom panel) in our ART (blue lines) and TNG simulations (red lines). \newtext{The SFR is estimated using young star particles with ages $<10\Myr$, while the shaded regions in the bottom panel indicate 16--84 interpercentile ranges of $\jzjc$.} Early on, star formation in ART is significantly more efficient and bursty than in TNG, and it exhibits two regimes: exponential ramping up at $z>7$ and steady decline at $z<7$. The transition point at $z\sim7$ corresponds to the formation of a thin gaseous disk forming stars with orbital circularities $\jzjc > 0.8$. In TNG, the $\Mstar$ and SFR evolution is significantly more gradual, and, although a coherently rotating galactic disk with $\jzjc>0$ also forms at $z\lesssim5$, the range of $\jzjc$ of young stars is significantly larger, implying a significantly thicker disk. These differences cannot be explained by the absence of AGN feedback in ART, as the resimulation of the TNG simulations without AGN feedback (red dashed line) produces qualitatively similar results to the original TNG run.}
\end{figure}

The differences in the gas density structure translate into the qualitative differences in the star formation histories. The top two panels of Figure~\ref{fig:sfh-jzjc} show the evolution of total stellar mass, $\Mstar$, and SFR of the main progenitor. With detailed modeling of cold ISM, the star formation becomes significantly more efficient and bursty. 

Early on, the SFR in ART is a factor of 10--30 higher than in TNG, leading to a significantly faster build-up of $\Mstar$. The shape of the early SFR history is also qualitatively different: in TNG it is rising monotonically, while in ART it rapidly rises until $z \sim 7$ and then steadily declines until the SFRs in the two runs become comparable in magnitude at $z \sim 3$ (although still significantly burstier in ART).  As we will discuss in the next section, this transition at $z \sim 7$ in ART corresponds to the formation of a galactic disk.

Apart from the average magnitude, the temporal variation of SFR is significantly higher in ART than in TNG. In TNG, the SFR evolves smoothly because local SFR depends only on the density of gas \citep{sh03} and, given the relatively smooth gas distribution (see Section~\ref{sec:results:dense-gas}), the total SFR is primarily set by the total amount of star-forming gas (defined using a relatively low-density threshold of $n \gtrsim 0.1\cc$) which changes smoothly. In contrast, in ART, star formation happens in relatively dense gas with low turbulence support (see Section~\ref{sec:methods:art}), and both the amount of such dense gas and the level of turbulence change rapidly, leading to a significant short-term variation of global SFR. 

Interestingly, just like the SFR magnitude, its variability also changes significantly before and after disk formation around $z\sim 7$. In a companion paper \citep{semenov24b}, we show that the short-term variability of the SFR is dominated by the local variations of $\epsff$ due to the chaotic changes in the ISM turbulence, which experience a qualitative change at the moment of disk formation. Before disk formation, these variations are more violent as they are driven by large-scale accretion flows and mergers. In contrast, after the disk formation, they are mainly caused by disk instabilities and stellar feedback driving the expansion--compression gas cycle in the ISM.

One of the important factors that potentially could contribute to the SFR being lower in TNG is the modeling of AGN feedback that is lacking in ART. However, we explicitly checked that AGN feedback is not efficient enough at these early times and low galaxy masses to explain the above differences. Specifically, we reran the TNG simulation with the AGN feedback turned off, and the results remained close to the original simulation, except for the later times ($z < 4$) or equivalently higher galaxy masses ($\Mstar > 10^9 \Msun$), where the run without AGN has a factor of 2 higher SFR and mildly larger $\Mstar$ (see the dashed red lines in the figure). It is worth noting however that, although the lack of AGN feedback in ART alone cannot explain the higher SFR and $\Mstar$, at $z < 7$ this galaxy in ART  becomes massive enough ($\Mstar > 10^9 \Msun$) for AGN to have a nonnegligible effect.
\newtext{It is worth noting, however, that supermassive black holes recently discovered in the early Universe appear to be significantly more massive relative to the stellar masses of their host galaxies than the local scaling relation would suggest \citep[e.g.,][]{pacucci23}, implying that their effect on early galaxy evolution could be substantial. All these questions warrant a separate investigation of AGN feedback in such early-forming galaxies.}

\subsection{Early Disk Formation}
\label{sec:results:disk-formation}

One of the striking results of the detailed ISM modeling is the early formation of an extended galactic disk (recall Figure~\ref{fig:maps}). As the bottom panel of Figure~\ref{fig:sfh-jzjc} shows, this disk forms in the ART simulation extremely early, at redshift $z \sim 6\text{--}7$.
This panel shows the orbital circularities\footnote{The orbital circularity is defined as the ratio of the $z$-component of the particle angular momentum, $j_z$, to the value at the circular orbit with the same total energy, $j_{\rm c}$. See Section 2.3 in \citet{semenov23a} for calculation details.} of stars \emph{at birth}, $\jzjc$, with values close to 0 and 1 corresponding to spherical and thin disk-like orbital configurations, respectively. In ART at $z > 7$, $\jzjc$ shows a large variation around 0 reflecting chaotic early evolution, while at $z \lesssim 7$, $\jzjc$ of most young stars becomes $\gtrsim 0.8$ reflecting the rapid formation of a relatively thin star-forming disk around $z \sim 6\text{--}7$. In TNG, the $\jzjc$ values also become predominantly positive around $z \sim 7\text{--}5$, but the scatter is large, reflecting a more puffed-up gas distribution albeit with a net rotation. It is worth recalling that this galaxy is not an average MW-mass galaxy progenitor, but rather an early-forming analog extracted from TNG50. However, while with the TNG model this early disk formed at $z \sim 3$ \citep{chandra23}, with a more detailed modeling of ISM physics, the same initial conditions lead to a 1.5 Gyr earlier disk formation at $z \sim 7$.

\begin{figure*}
\centering
\includegraphics[width=0.5\textwidth]{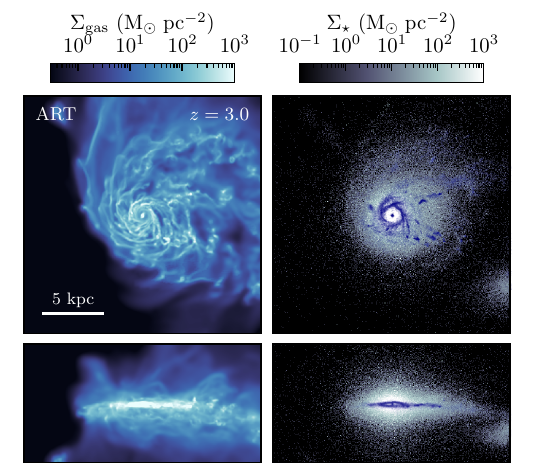}%
\includegraphics[width=0.5\textwidth]{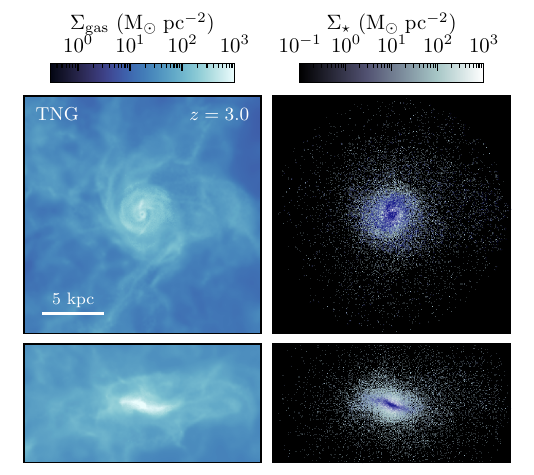}\\
\caption{\label{fig:disk-maps} Face-on and edge-on views of the gas and stellar disk in our ART (left set of panels) and TNG simulations (right set of panels) at $z=3$. Blue points in the stellar maps indicate the location of stars younger than $100\Myr$. With a more detailed ISM modeling, the disk in ART is significantly thinner and exhibits a richer structure than that in TNG.}
\end{figure*}

\begin{figure*}
\centering
\includegraphics[width=0.5\textwidth]{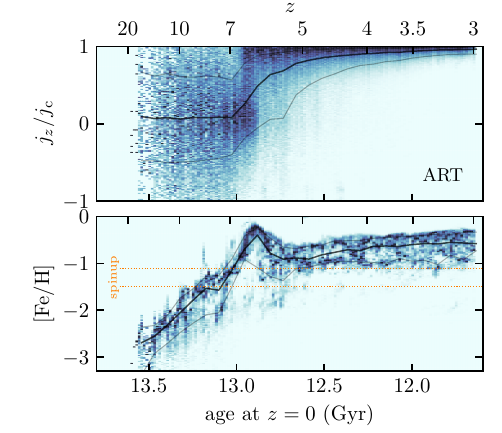}%
\includegraphics[width=0.5\textwidth]{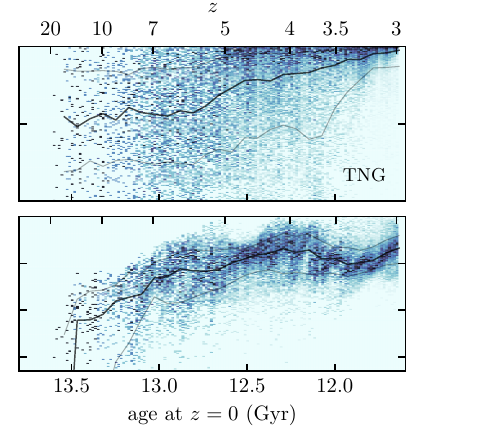}
\caption{\label{fig:disk-FeH-jzjc} Redshift $z=3$ distributions of stellar iron abundances, [Fe/H], and orbital circularities, $\jzjc$, as a function of age in our ART (left) and TNG (right) simulations. For convenience, the stellar ages are calculated for $z=0$. The black lines show the median and $16^{\rm th}\text{--}84^{\rm th}$ interpercentile range (i.e., $\sim 1 \sigma$ variation). The dotted orange lines indicate the range of [Fe/H] over which the disk spinup feature occurs in the distribution of stellar tangential velocities in ART (see Figure~\ref{fig:FeH-pdf-spinup} below). The abrupt emergence of the star-forming gas disk at $z \sim 6\text{--}7$ is clearly visible in the distribution of stellar $\jzjc$ at $z=3$, indicating that the star-forming disk is sufficiently stable to build up the stellar disk steadily. In TNG, the stellar disk at $z=3$ is significantly less prominent. }
\end{figure*}

Note that Figure~\ref{fig:sfh-jzjc} shows stellar circularities \emph{at birth}, implying that it is a star-forming gas disk that forms at $z \sim 6\text{--}7$. By itself, this does not necessarily imply a buildup of a stable stellar disk as early gas disks can be transient or change orientation on short timescales \citep[e.g.,][]{meng-gnedin21,kretschmer22}. However, in Figures~\ref{fig:disk-maps} and \ref{fig:disk-FeH-jzjc}, we show that in our simulation the gas disk is sufficiently stable, and a stellar disk is steadily built up after the formation of the star-forming gas disk.

Figure~\ref{fig:disk-maps} shows the spatial distribution of gas and stars at $z=3$, viewed face-on and edge-on. Young stellar particles formed within 100 Myr are also highlighted in blue. In ART, the young stellar disk is relatively thin, reflecting the thin star-forming gas disk. The vertical extent of the entire stellar population is, however, thicker due to the contribution of stars formed before a disk was established (the population dubbed ``Aurora'' by \citealt{bk22}, or ``proto-Galaxy'' by \citealt{conroy22}), as well as thickening of the population of stars formed shortly after gas disk formation (see below). In contrast, although in TNG a disk-like component is also clearly visible in gas and young stars, it is significantly thicker, consistent with the large scatter of $\jzjc$ in Figure~\ref{fig:sfh-jzjc}.

\begin{figure}
\centering
\includegraphics[width=\columnwidth]{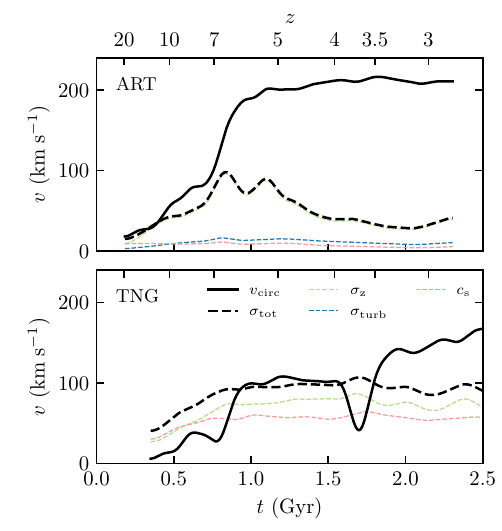}
\caption{\label{fig:vcirc-sigma} The evolution of characteristic velocities competing in disk formation: the circular velocity, $\vcirc = \sqrt{GM_{\rm tot}/R}$, at the radius of full rotational support for the dense gas with $n > 0.3\cc$ (solid line), and the total velocity dispersion of gas ($\stot$; dashed lines). The colored dashed lines also show the contributions to $\stot$ from the resolved vertical velocity dispersion ($\sigma_z$; green), subgrid turbulent velocity ($\st$, blue), and thermal sound speed ($\cs$; red). Note that $\stot$ in this plot is different from the one used in the definition of the subgrid $\avir$ (Equation~\ref{eq:avir}), as the latter is computed locally and does not include $\sigma_z$. A thin disk can form only when $\vcirc$ reaches above the total velocity dispersion and the characteristic velocity of SN-driven flows ($\sim 100\kms$). This happens significantly earlier in ART owing to the more efficient accumulation of mass in the halo center (leading to higher $\vcirc$) and weaker thermal support combined with efficient dissipation of turbulent motions (leading to lower $\stot$).}
\end{figure}

The top panels of Figure~\ref{fig:disk-FeH-jzjc} show the $z=3$ distribution of stellar $\jzjc$ as a function of age. Overall, this distribution qualitatively mirrors the distribution of $\jzjc$ \emph{at birth} (the bottom panel in Figure~\ref{fig:sfh-jzjc}), which implies that the gaseous disk in ART does remain in place and gradually builds up the stellar disk without drastic changes in orientation or encounters with destructive mergers between $z\sim7$ and 3 \citep[see also][]{yu22}. In particular, the sharp increase of $\jzjc$ at $z \sim 6\text{--}7$ is clearly visible, reflecting the signature of disk spinup in the $z=3$ stellar kinematics. Note, however, that the scatter of $z=3$ values of $\jzjc$ for stellar particles with ages $\sim$12.4--12.9 Gyr (i.e., shortly after disk formation) is wider than that of the $\jzjc$ values at birth. This difference reflects the thickening of the early stellar disk as a result of the aforementioned variation in the gas disk orientation and secular dynamic processes.

For the comparison with the local Galactic archaeology data (see the next section), the chemical evolution of the stellar population is crucial as ages are not observable directly. The lower panels of Figure~\ref{fig:disk-FeH-jzjc} show the distribution of stellar metallicities, [Fe/H], as a function of age. In ART, [Fe/H] increases rapidly and peaks at $z \lesssim 7$, which happens around the end of disk formation. This peak in [Fe/H] coincides with a peak in the SFR (see Figure~\ref{fig:sfh-jzjc}) which leads to rapid enrichment of the ISM, while the subsequent drop and steady increase of [Fe/H] reflects mixing in of the metal-poor material right after disk formation and steadily declining SFR in the disk. 
In TNG, the evolution of [Fe/H] is significantly smoother. Interestingly, it also exhibits a larger scatter than in ART. This is a consequence of the adopted star formation model with a relatively low-density threshold, which results in a more distributed star formation (recall Figure~\ref{fig:disk-maps}) and therefore probes the available metallicity gradient of the gas disk more uniformly.

The quantitative insights into why the explicit modeling of cold ISM leads to significantly earlier disk formation are provided in Figure~\ref{fig:vcirc-sigma}. The solid line shows the evolution of the circular velocity, $\vcirc = \sqrt{GM_{\rm tot}/R}$, at the circularization radius of the dense gas with $n > 0.3\cc$, i.e., the radius at which this gas would be fully rotationally supported given its total angular momentum. A thin disk can form only when this velocity becomes larger than the velocity dispersion of the gas (shown with the dashed lines) and a characteristic velocity of feedback-driven flows that remove gas from the region. The exact value of the latter depends on the details of star formation feedback modeling, but for typical energetics of SNe it is of the order of $\sim 100\kms$ \citep[e.g.,][]{dekel86}. 

In ART, $\vcirc$ gradually rises prior to disk spinup at $z\sim7\text{--}6$ and then increases sharply as it reaches $\sim 100\kms$, when the disk forms. The velocity dispersion during disk formation is dominated by turbulent motions on resolved scales, $\sigma_{z} \lesssim 80 \kms$, with only a small contribution of thermal and subgrid turbulent velocities, both at $\sim 10\kms$. The resolved turbulence decays down to $\sigma_{z} \lesssim 40 \kms$ over the next $\sim 0.5\Gyr$ as the disk settles.

In TNG, in contrast, $\vcirc$ increases more gradually and rises significantly above $\sim 100\kms$ only at $z \lesssim 3.5$. The total velocity dispersion also remains high, $\sim 100\kms$, due to both large resolved velocity dispersion and the thermal energy associated with the effective equation of state. This leads to a significantly thicker disk in TNG at $z = 3$.

Overall, the differences in the evolution of $\vcirc$ arise from the differences in how efficiently mass can be accumulated in the center of the halo. 
The early disk formation in the ART simulation is therefore linked to the more rapid growth of the stellar mass at early times. Previous numerical studies showed that galactic disks tend to form when galaxies become more massive than $\sim 10^9\Msun$ \citep[e.g.,][]{el-badry18,dekel20,pillepich19,dillamore24,semenov23b}, although the exact physical mechanism of disk spinup remains a subject of debate \citep[e.g.,][]{stern19,dekel20,hopkins23disk,semenov23b}. As Figure~\ref{fig:sfh-jzjc} shows, the disk forms in both ART and TNG when the galaxy reaches such a mass of $\Mstar \sim 2\times10^9\Msun$. With more efficient star formation, the galaxy in the ART simulation crosses this mass threshold earlier, at $z \sim 7$, leading to earlier disk formation than in TNG.

\subsection{Comparison with Milky Way Data}
\label{sec:results:mw-data}

\begin{figure}
\centering
\includegraphics[width=\columnwidth]{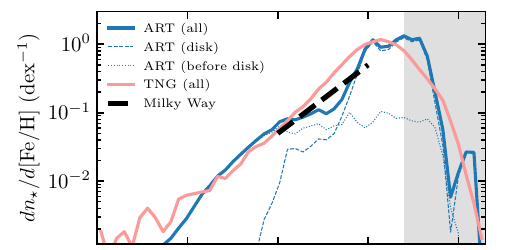}\\
\includegraphics[width=\columnwidth]{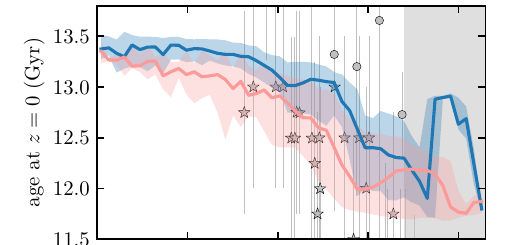}\\
\includegraphics[width=\columnwidth]{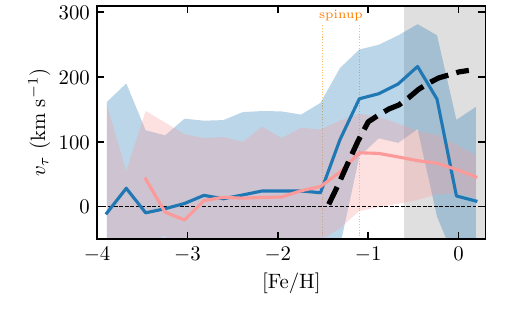}
\caption{\label{fig:FeH-pdf-spinup} Comparison of the stellar disks in our simulations at $z=3$ with the Milky Way data at $z=0$: the distribution of stellar [Fe/H] (top panel), age--metallicity relation (middle panel), and the dependence of the galactocentric tangential velocity, $\vt$, on [Fe/H] (bottom panel). The solid line and shaded region in the middle and bottom panels show the median and $16^{\rm th}\text{--}84^{\rm th}$ interpercentile range (i.e., $\sim 1 \sigma$ variation), respectively. Only the comparison at low metallicity is sensible, as higher-metallicity stars form at later times and therefore are not fully sampled in our simulation; for this reason, we gray out the values of [Fe/H] $>-0.6$ corresponding to the median metallicity at $z=3$ in our simulations. The metallicity PDF at $-2 < \text{[Fe/H]} < -1$ roughly follows a linear slope consistent with the results of \citet{rix22} and \citet{bk23}, although the ART simulation shows a feature within this range caused by disk formation (see the text). The age--metallicity relation of both simulations is also consistent with the MW data for both the population of low-metallicity globular clusters \citep[stars;][]{bk24} and metal-poor disk stars \citep[circles;][]{woody24}. Stellar $\vt$ also shows a sharp spinup feature at $-1.5 < \text{[Fe/H]} < -1$, also consistent with the findings reported by \citet{bk22}, while in TNG this feature at $z=3$ is less prominent.  }
\end{figure}

The direct comparison of our simulation results with the MW data is impossible because the simulation was not run to $z=0$ due to the computational expense. However, it is still insightful to compare the properties of the high-redshift disks in our simulation with the distribution of MW's low-metallicity stars. Most of these stars form early, and their metallicities and kinematics imprint useful information about these early stages of galaxy formation. 
In particular, Figure~\ref{fig:FeH-pdf-spinup} compares the probability density function (PDF) of [Fe/H], age--metallicity relation, and the tangential velocity of stars as a function of [Fe/H] with the MW measurements. The high-metallicity region with [Fe/H] $>-0.6$ corresponding to the median gas metallicity at $z=3$ is grayed out because, by $z=0$, it will be populated predominantly by stars formed at $z<3$.

\begin{figure}
\centering
\includegraphics[width=\columnwidth]{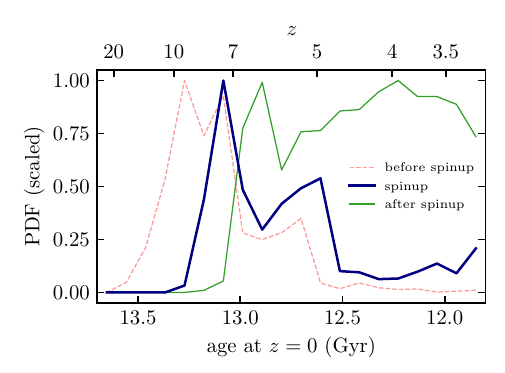}
\caption{\label{fig:spinup-age} Distribution of stellar ages in three metallicity bins corresponding to [Fe/H] before, within, and after the range over which the disk spinup feature occurs in the bottom panel of Figure~\ref{fig:FeH-pdf-spinup} (specifically, using [Fe/H] = $-1.5$ and $-1.1$ thresholds indicated with the dotted vertical lines in that figure).  To factor out the drastic difference in stellar mass between these populations, the distributions are scaled such that their maxima equal one. The low (before spinup) and high (after spinup) metalicity populations correspond to stellar particles with ages $>13$ and $<13$ Gyr respectively, while the intermediate population corresponding to the spinup feature exhibits a relatively wide distribution of ages (see the text).}
\end{figure}

As the top panel shows, the average slope of the metallicity PDF at $-2 \lesssim \text{[Fe/H]} \lesssim -1$ roughly follows the linear relation found in the MW \citep{rix22,bk23}, implying that both TNG and ART models lead to reasonably realistic early enrichment histories. Using the TNG simulation ran to $z \sim 0$, we have confirmed that the PDF slope at these low metallicities remains close to linear and only the PDF portion at $\text{[Fe/H]} > -1$ grows substantially at $z<3$. The PDF in ART exhibits a dip around $\text{[Fe/H]} \sim -1.5$, which, in this case, imprints the onset of disk formation, as is evident from the comparison of PDFs for stars formed before and after the disk was formed (thin dotted and dashed lines, respectively). A similar feature is reported in the MW data, albeit at somewhat higher metallicity, $\text{[Fe/H]} \sim -1$ (see the references above).

Similarly, the comparison of the age--metallicity relations in the middle panel shows that both TNG and ART simulations are consistent with the MW observations at these early times. This is mainly because the uncertainties in ages within and between the models are significant, $\sim$1 Gyr, which is comparable with the age of the universe at these early times. Interestingly, both ART and TNG runs exhibit rather smooth relations at low metallicities; in particular, the sharp increase of [Fe/H] at age $\approx$12.9 Gyr in ART (recall Figure~\ref{fig:disk-FeH-jzjc}) is not visible when the age distribution is binned as a function of [Fe/H]. This increase is manifested in the feature at near-solar metallicity, however, by $z=0$ its contribution into the total population of metal-rich stars becomes overpowered by the stars formed in the metal-rich disk at late times. 

The bottom panel shows the relation between the galactocentric tangential velocity of stars, $\vt$, as a function of their metallicity. The dashed line shows the first discovery of the sharp increase of spinup $\vt$ in the low-metallicity population of the MW stars corresponding to the spinup of the MW disk reported by \citet[][see also \citealt{conroy22}]{bk22}. As the figure shows, in ART such a spinup feature is clearly in place by $z=3$, while TNG shows a significantly lower amplitude of median $\vt$. If the stellar disk did not encounter destructive mergers at later times, such a feature could potentially survive, indicating that it could reflect the disk formation as early as $z\sim 6\text{--}7$, as the case in our simulation.

\subsection{Timing of Disk Spinup}
\label{sec:results:disk-timing}

Given that the early stellar and gas disks during and shortly after the formation are rather weak in terms of the common disk definitions, such as thickness-to-radius ratio and rotational support, the timing of disk formation can also be ambiguous and is subject to the disk definition. For example, while the gas disk in ART has rather low $V/\sigma \sim 1\text{--}2$ at $z\sim 7\text{--}5$ (Figure~\ref{fig:vcirc-sigma}), the stellar $\jzjc$ at $z=3$ show a clear signature of disk spinup at $z\sim 6\text{--}7$ (Figure~\ref{fig:disk-FeH-jzjc}) consistent with the formation of a star-forming gas disk at the same redshifts (Figure~\ref{fig:sfh-jzjc}). To avoid ambiguity, we define the disk formation stage as the time over which the spinup feature in the distributions of $\jzjc$ is built up. From Figures~\ref{fig:sfh-jzjc} and \ref{fig:disk-FeH-jzjc}, this stage spans between the stellar ages $\sim13$ and 12.8 Gyr,  implying disk formation on a $\sim$0.2 Gyr timescale.

The disk formation time implied by the observed [Fe/H]--$\vt$ relation (Figure~\ref{fig:FeH-pdf-spinup}) can differ from these values as stars with a range of ages can contribute to the same [Fe/H] bin due to the scatter in the age metallicity relation.
To investigate this effect, Figure~\ref{fig:spinup-age} shows the distributions of stellar ages before ([Fe/H] $<-1.5$), during ($-1.5<$ [Fe/H] $<-1.1$), and after ([Fe/H] $>-1.1$) the spinup selected based on the metallicity cuts corresponding to the sharp increase of $\vt$ (see the dotted orange lines in Figure~\ref{fig:FeH-pdf-spinup}). 
The star particles with [Fe/H] $<-1.5$ and $>-1.1$ are relatively cleanly separated into the population formed at $z > 7$ and $z < 7$, or ages $> 13$ and $<13$ Gyr, respectively. The intermediate population corresponding to the spinup feature, however, shows a relatively broad distribution of ages, from 13.2 to 12.5 Gyr. Na\"{i}vely, this could be interpreted as disk formation on a timescale of 0.7 Gyr. However, this range is significantly broader than the aforementioned definition based on the stellar ages: $\sim13$--12.8 Gyr, implying a significantly more abrupt disk formation on 0.2 Gyr timescale. 

The wider range of spinup ages in the [Fe/H]--$\vt$ space originates from the scatter of [Fe/H] at a fixed age, as evident from Figure~\ref{fig:disk-FeH-jzjc} (see the dotted orange lines showing the same [Fe/H] cuts corresponding to the spinup feature). The bump and tail of spinup stars with ages $<12.8\Gyr$ in Figure~\ref{fig:spinup-age} correspond to the low-metallicity end ([Fe/H] $<-1.1$) of the stellar population formed in the disk at late times. Also interestingly, the narrow peak of the oldest subpopulation of these spinup stars with ages $\sim13$--$13.2$ Gyr was formed shortly \emph{before} a stable disk was established, not during the disk formation. This is because the abrupt disk formation coincides with the peak of star formation and metal enrichment (recall Figure~\ref{fig:sfh-jzjc}), so that the stars with $\jzjc \sim 1$ and large $\vt$ formed during this peak contribute significantly at [Fe/H] $\gtrsim -1.1$, while the stars with [Fe/H] $\lesssim -1.1$ reflect the transition to the population formed before the disk.

Thus, although the spinup feature observable in the [Fe/H]-space corresponds to the sharp spinup in the age-space, the scatter and shape of the age--metallicity relation can lead to a biased estimate of disk formation in terms of both the timing and duration: 13.2--12.5 Gyr based on [Fe/H] vs. 13--12.8 Gyr based on ages. These differences, however, are comparable to the uncertainties in the age estimates at these low metallicities. It is nevertheless interesting that the timescale of disk formation implied by [Fe/H] is consistent with the observational estimates ($\sim 1$ Gyr; e.g., \citealt{bk24}) while our simulation predicts that disk formation is in fact significantly more abrupt ($\sim 0.2$ Gyr).

\subsection{Comparison with Observed Early Disk Galaxies}
\label{sec:results:jwst-alma}

\begin{figure*}
\centering
\includegraphics[width=0.5\textwidth]{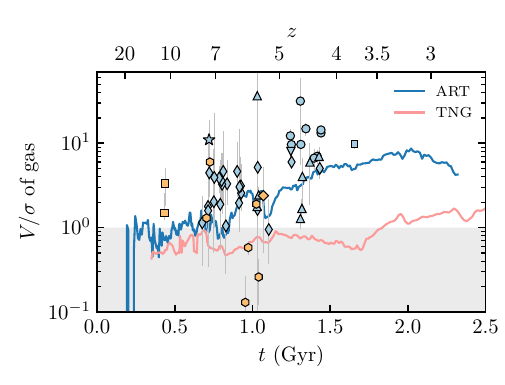}%
\includegraphics[width=0.5\textwidth]{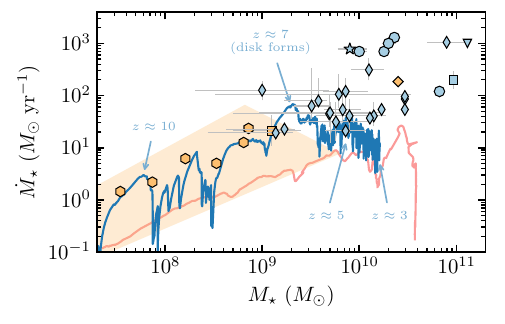}
\caption{\label{fig:M-SFR-vrot-sigma} Comparison of our simulation results with the observed early disk galaxies at $z>4$: rotational support of the gaseous disk ($V/\sigma$; left) and the $\Mstar$--SFR relation (i.e., the main sequence; right). The blue and orange markers show the galaxies with estimated $V/\sigma$ detected in ALMA (blue colors; circles from \citealt{rizzo21}; upward triangles from \citealt{jones21}; downward triangles from \citealt{lelli21}; diamonds from \citealt{parlanti23}; square from \citealt{amvrosiadis24}, and star from \citealt{rowland24}) and \emph{JWST} (orange color; diamond from \citealt{nelson23}; hexagons from \citealt{degraaff24}; square from \citealt{xu24}), respectively. The orange polygon in the right panel also indicates the typical range of $\Mstar$ and SFR derived for the galaxies at $z\sim 10$ observed with \emph{JWST} (compilation of observational results from Figure 6 in \citealt{kravtsov-belokurov24} that includes the data from \citealt{atek23,roberts-borsani23,franco23,morishita23,morishita24}). A few redshift values are also indicated for the ART run in the right panel. The ART simulation is consistent with some of the observed early disk galaxies both in terms of rotational support and their location on the main sequence implying that some of these galaxies might be progenitors of early-forming MW analogs. The variation of SFR at $z>7$ in ART is also consistent with the observed scatter.}
\end{figure*}

The fact that detailed turbulence-based ISM modeling leads to efficient bursty star formation and early disk formation is enticing in view of the recent \emph{JWST} and ALMA  discoveries of large abundances of early bright galaxies and dynamically cold gaseous disks (see the references in the Introduction). Figure~\ref{fig:M-SFR-vrot-sigma} compares the results of our simulations with observational estimates of the gaseous disk rotational support ($V/\sigma$; left panel), and the relation between $\Mstar$ and SFR, i.e., the main sequence (right panel). In simulations, we calculate the rotational velocity of the gas disk as $V = \sqrt{2\, E_{\tau,\rm gas}/M_{\rm gas}}$, where $E_{\tau,\rm gas}$ and $M_{\rm gas}$ are the tangential component of kinetic energy and the total mass of gas denser than $n>0.3\cc$,\footnote{The estimate based on the circular velocity, $\sqrt{G\,M_{\rm tot}/r}$, at the circularization radius of gas produces a similar result, albeit with a larger scatter at early times. \newtext{We have also checked that the results in ART remain almost unchanged when we use only cold gas ($T < 2000\K$) instead of all gas denser than $n>0.3\cc$. For consistency with TNG, where cold gas is not available, we present the results for the gas selected by the density threshold.}} while the total vertical velocity dispersion, $\sigma$, includes the contributions from the resolved and thermal motions and, in the ART case, subgrid turbulence.  

As the left panel shows, the gas disk in the ART simulation has $V/\sigma \sim 3\text{--}10$ which is consistent with the rotational support of some of the early ALMA and \emph{JWST} disks at $z \sim 4\text{--}6$. As the right panel shows, our simulation is also consistent with some of these galaxies in terms of the total stellar mass and the SFR. In particular, GN-z11 (orange square; \citealt{xu24}) and the disk galaxies from the JADES sample (orange hexagons; \citealt{degraaff24}) are consistent with our simulated MW progenitor before the formation of the thin disk at $z\approx 7$. Many of the ALMA cold disks from the \citet{parlanti23} sample at $z=5\text{--}7$ (blue diamonds) are consistent with being MW-like progenitors with early disk formation. Many of the observed high-redshift disks are, however, too massive ($\Mstar > 2\times 10^{10}\Msun$) and/or actively star-forming ($\SFR > 100\Msunyr$; e.g., \citealt{lelli21,rizzo21,nelson23,amvrosiadis24,rowland24}), implying that these galaxies are likely progenitors of more massive early-type galaxies in centers of groups and clusters.

The orange polygon in the right panel shows the range of $\Mstar$ and $\SFR$ derived for \emph{JWST} galaxies at $z\sim10$. The SFRs of these galaxies are systematically above that in TNG.\footnote{Note, however, that the observed range corresponds to the brightest galaxies at these early times, so the actual range of SFR can extend to smaller values.} The ART galaxy, on the other hand, explores almost the entire range of the observed SFRs as it evolves, implying that the early, pre-disk SFR burstiness is consistent with the observed scatter. In a companion paper \citep{semenov24b}, we show that this variation is sufficient to explain the abundance of early UV-bright galaxies observed with \emph{JWST}, and most of this variation results from variations in local star formation efficiency per freefall time ($\epsff$; see Equation~\ref{eq:rhosfr}) caused by variations in the ISM turbulence. This can be one of the reasons why many of the existing galaxy simulations underproduce the abundance of early bright galaxies without modeling such effects.

\section{Discussion}
\label{sec:discussion} 

In our simulation of an MW-like progenitor with detailed ISM modeling, a stable star-forming disk forms as early as $z \sim 6\text{--}7$ (Section~\ref{sec:results:disk-formation}). The disk at these early times differs from typical star-forming disks in the present-day universe. For example, at $z \sim 3$ the contribution of the pre-disk stellar population is still significant, so that the galaxy exhibits a significant spheroid contribution (Figure~\ref{fig:disk-maps}). The rotational support of the gas disk is also rather weak: $V/\sigma \sim 3$ shortly after the formation, and increasing to $V/\sigma \sim 5\text{--}8$ at $z < 5$, reflecting high level of ISM turbulence at these early times (Figure~\ref{fig:vcirc-sigma}). These numbers are however consistent with the observations of disk galaxies at $z > 4$ with \emph{JWST} and ALMA (Figure~\ref{fig:M-SFR-vrot-sigma}).

The ICs for this simulation were selected from the TNG50 simulation as one of the closest MW analogs in terms of the chemo-kinematic structure of the stellar disk at $z=0$ \citep[subfind halo ID {\tt 519311}; see Section~\ref{sec:methods:ics} and][]{chandra23}. Interestingly, in our resimulation of this galaxy with detailed ISM modeling, some of the key properties of the low-metallicity disk are already in place early on. In particular, the sharp increase in the tangential velocity of stars at metallicities $-1.5 <$ [Fe/H] $< -1$, is consistent with the MW data that reflects the MW disk formation \citep[Figure~\ref{fig:FeH-pdf-spinup};][]{bk22,conroy22}. In our simulations, this feature is established at $z \sim 6\text{--}7$ (Section~\ref{sec:results:disk-timing}). 

One of the key reasons for the earlier disk formation in our enhanced-physics simulation is the more efficient accumulation of the baryonic mass in the galaxy early on. Both the original TNG galaxy and our resimulation form a stable star-forming disk, when the galaxy stellar mass reaches $\sim 2 \times 10^{9}\Msun$, consistent with previous findings \citep[see Section~\ref{sec:results:disk-formation} and ][]{el-badry18,dekel20,pillepich19,dillamore24,semenov23a,semenov23b}. With detailed ISM modeling, our simulated galaxy forms stars more efficiently and reaches this mass threshold earlier than in TNG (recall Figure~\ref{fig:sfh-jzjc}). Given that, in both simulations, the host dark matter halo mass evolves similarly, the enhanced-physics run predicts higher $\Mstar/M_{\rm halo}$ at high redshift than TNG. Note, however, that we focus on the evolution of an atypical (early-forming) halo, and therefore, a detailed comparison with the observational constraints and empirical and semianalytic models requires a more uniform sample of simulated galaxies, which we leave to a future study.

\newtext{The more efficient early star formation that leads to the early disk formation in our simulation can be viewed as stellar feedback being on average less efficient in suppressing global star formation at these early times compared to the TNG model. Note, however, that stellar feedback still needs to be sufficiently efficient to ensure the survival of the early disk via its effects on disk stability and angular momentum (see references in the Introduction). Therefore, the timing of disk formation and its imprints in the present-day chemo-kinematic structure of galaxies can be used as a stringent constraint of star formation and feedback modeling \citep[e.g.,][]{yu21,yu22,bk22,bk23,khoperskov22c,mccluskey23,semenov23a,semenov23b}. For example, as we show in the companion paper, using common star formation prescriptions with local star formation efficiency per freefall time fixed at high values can hinder the formation of galactic disks \citep{semenov24b}.}

Considered together, our results suggest that the MW disk could spin up very early, and some of the observed early disks at $z > 4$ could be progenitors of present-day MW-like systems. If the MW's disk has indeed formed that early, this would require that the kinematic distribution of the stars formed at these early times does not evolve substantially over the remaining history of the universe due to secular processes, like stellar migration, and catastrophic events, like galaxy mergers. While numerical simulations suggest that stellar disks might indeed be resilient to the former long-term evolution \citep[e.g.,][]{yu22}, galaxy mergers can indeed be destructive posing stringent constraints on the late mass accretion history for such early-forming disks.

As was shown by \citet{sotillo-ramos22}, a significant fraction of disk galaxies in TNG can survive mergers. This is also the case for the TNG50 galaxy that we study here, which undergoes a series of significant mergers at $z\sim3$ and $0.7$ that shape the final structure of the stellar disk in the [Fe/H] vs. $\jzjc$ space, which resembles the MW data \citep{chandra23}. By resimulating our ART run at lower resolution past $z \sim 0.7$ we find that, with the physics adopted in our simulation, this is no longer the case: the first merger destroys the disk formed at $z \sim 7$, shifting the spinup feature to significantly higher [Fe/H] than the MW data suggests (recall Figure~\ref{fig:FeH-pdf-spinup}), while the later merger quenches the galaxy completely.

This drastic difference between ART and TNG results indicates that the resilience of stellar disks to mergers is strongly sensitive to the modeling of the ISM, star formation, and feedback. There are multiple ways how such details can play a role. First, by making star formation more efficient early on, more gas is converted into stars in the ART run, making mergers more gas-poor, affecting disk survivability and reformation after the merger. Second, the development of instabilities in the gas disk and its response to mergers depend on the ISM structure and its effective EoS which are significantly different in simulations with an imposed EoS and explicit modeling of multiphase ISM structure formation (see Section~\ref{sec:results:dense-gas} and Figures~\ref{fig:n-T}--\ref{fig:n-sigma}). Third, mergers drive a significant fraction of gas mass into the center, making the subsequent evolution dependent on the details of the AGN feedback modeling, which is absent in our ART runs. Finally, we find that the modeling of a cold ISM substantially changes the timing and orbital parameters of mergers \citep[see also][]{agora24}. All these effects of ISM modeling on mergers require a further detailed study, which we leave to future work.

The observational estimates of the total mass of MW stars formed before its disk was in place (i.e., the ``Aurora'' or ``proto-Galaxy'' population) suggest that no such destructive mergers should have happened in the MW's history. Indeed, this mass is rather low ($\sim 10^8 \Msun$ in \citealt{conroy22} and $\sim 10^9 \Msun$ in \citealt{bk23}), which is close to the stellar mass threshold at which galaxies tend to form disks \citep[e.g.,][]{el-badry18,dekel20,pillepich19,dillamore24,semenov23b}. If a destructive merger occurred, the total mass of non-disky low-metallicity stars in the MW would be higher as it would include both the contribution of stars that were formed before any disk was in place and the stars formed in a disk that was destroyed by the merger. The comparison of MW's data and cosmological simulation results also suggests that the MW could indeed experience a merger after the disk formation, but its disk survived through this last major merger \citep[][]{fattahi19,dillamore22,dillamore24}.

Thus, whether the early disk survives or not also strongly depends on the full mass assembly history and the presence of significant mergers. The fact that the disk does not survive in the ART simulation can be an artifact of the particular choice of the ICs. Indeed, halo {\tt 519311} experiences more significant late mergers compared to some other early-forming MW-like objects from TNG50 \citep[see, e.g., Figure~8 in][]{chandra23}. This warrants a more detailed exploration of ICs for other early-forming galaxies with detailed ISM modeling, which we leave to a future study.

\section{Summary and Conclusions}
\label{sec:summary}

We have investigated the early evolution and galactic disk formation in a cosmological zoom-in simulation of a close MW-like analog, incorporating detailed modeling of cold ISM formation coupled with on-the-fly transfer of UV radiation and turbulence-regulated star formation and stellar feedback. For the initial conditions of our simulation, we have extracted one of the early-forming MW-mass disk galaxies from TNG50 that closely resembles the MW in terms of the early formation of the stellar disk and its chemo-kinematic structure at $z=0$ \citep{semenov23a,chandra23}. To assess the effects of detailed ISM, star formation, and feedback modeling, we compare the results of our new resimulation with the original TNG model. Our main findings are summarized as follows:

\begin{enumerate}
    \item With the explicit modeling of cooling and heating, the ISM at densities $n \gtrsim 0.1\cc$ develops a two-phase structure with characteristic temperatures of $T \sim 10^4\K$ (warm neutral medium and resolved \ion{H}{2} regions) and $T < 100\K$ (cold neutral medium). In contrast, for TNG, the effective equation of state imposed at these densities corresponds to significantly higher temperatures, $T \sim 10^4\text{--}10^5\K$, thereby preventing the formation of dense structures (Figures~\ref{fig:maps}--\ref{fig:n-sigma}).

    \item As a result of the difference in gas structure and star formation and feedback modeling, the new simulation exhibits a significantly earlier and bursty star formation history. The global SFR exhibits a steep rise at $z>7$ and then a steady decline at $z<7$. In TNG, in contrast, the early SFR is very smooth and monotonically rising (the top two panels of Figure~\ref{fig:sfh-jzjc}).

    \item The transition between these two star formation regimes in our enhanced-physics simulation corresponds to the formation of a stable star-forming gas disk at $z\sim6\text{--}7$. The initial conditions for this run were selected from TNG50 such that the disk also forms relatively early, but with the TNG model, this happens significantly later, at $z \sim 3$. In both cases, the gas disk forms when the galaxy stellar mass reaches $\sim 2 \times 10^9 \Msun$, consistent with previous findings in the literature (the bottom panel of Figure~\ref{fig:sfh-jzjc}).

    \item The chemo-kinematic properties of this early-formed disk are consistent with the low-metallicity ([Fe/H] $\lesssim -0.6$) population of the MW stars in terms of the stellar metallicity distribution, age--metallicity relation, and the sharp increase of rotational velocities at $-1.5 \lesssim$ [Fe/H]$\lesssim -1$, indicative of the early MW's disk spinup (Figure~\ref{fig:FeH-pdf-spinup}). The galaxy in TNG also develops a rotating disk-like system by $z \approx 3$, which is however significantly thicker than in the simulation with detailed ISM modeling (the bottom panel of Figure~\ref{fig:sfh-jzjc} and Figure~\ref{fig:disk-maps}).

    \item The disk in the enhanced-physics simulation forms very abruptly, within $\sim0.2$ Gyr. This is significantly shorter than the estimate based on the age distribution of stars constituting the above spinup feature suggests, $\sim0.7$ Gyr. The latter estimate is biased due to the scatter of stellar metallicities at a fixed age resulting in a low-metallicity tail of stars formed in the disk contributing to the spinup feature. The difference is however smaller than current uncertainties in the observational age estimates (Section~\ref{sec:results:disk-timing}).

    \item The rotational support of the gas disk at $z\sim4\text{--}7$ is consistent with the observations of early disk galaxies with \emph{JWST} and ALMA in this redshift range. The location of the galaxy in the main sequence is also consistent with the low mass ($\Mstar < 2\times 10^{10}\Msun$) and low SFR ($\SFR < 100\Msunyr$) end of this galaxy sample. This indicates that at least some of these observed early disk galaxies can be progenitors of MW-like systems (Figure~\ref{fig:M-SFR-vrot-sigma}).

    \item The variation of the global SFR before the disk formation is consistent with the observed scatter, while the TNG model predicts the evolution along the main sequence without significant SFR variations and with a magnitude $\sim$10 times lower than the average values derived observationally (Figure~\ref{fig:M-SFR-vrot-sigma}, right panel).
\end{enumerate}

Our results demonstrate that modeling of turbulent cold ISM, and turbulence-regulated star formation and feedback can both produce highly variable early SFR---which helps to reconcile early galaxy evolution with the high abundance of UV-bright galaxies observed by \emph{JWST} at high redshifts---while at the same time enable early disk formation, without destroying it with excessively efficient stellar feedback.

We find that the early disk in our simulation does not survive to $z=0$ due to a subsequent destructive merger. This can be either an artifact of the specific choice of the ICs or a generic consequence of more efficient early star formation (see Section~\ref{sec:discussion}). Both these possibilities warrant future investigation in a larger sample of early-forming disk galaxies with a range of mass assembly histories. 

In a companion paper \citep{semenov24b}, we use our new simulation to investigate the role of ISM turbulence in the early SFR burstiness and disk formation in more detail. Forward-modeling of the small-scale turbulence and local star formation efficiencies on tens of parsec scales, make these simulations particularly useful as they allow us to investigate these properties in the early galaxies where observational estimates are sparse or lacking altogether.

\section*{Acknowledgements}
We are deeply grateful to Oscar Agertz, Renyue Cen, and Andrey Kravtsov for the insightful discussions and comments. 
The analyses presented in this paper were performed on the FASRC Cannon cluster supported by the FAS Division of Science Research Computing Group at Harvard University.
Support for V.S. was provided by Harvard University through the Institute for Theory and Computation Fellowship. 
Analyses presented in this paper were greatly aided by the following free software packages: {\tt yt} \citep{yt}, {\tt NumPy} \citep{numpy_ndarray}, {\tt SciPy} \citep{scipy}, {\tt Matplotlib} \citep{matplotlib}, and \href{https://github.com/}{GitHub}. We have also used the Astrophysics Data Service (\href{http://adsabs.harvard.edu/abstract_service.html}{ADS}) and \href{https://arxiv.org}{arXiv} preprint repository extensively during this project and writing of the paper.

\appendix

\section{Resolution}
\label{app:resolution}

\subsection{ART vs TNG Comparison}
\label{app:resolution:comparison}

\begin{figure}
\centering
\includegraphics[width=\columnwidth]{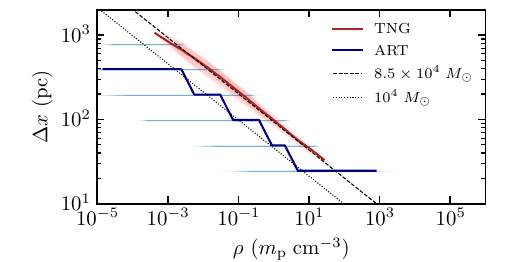}\\
\includegraphics[width=\columnwidth]{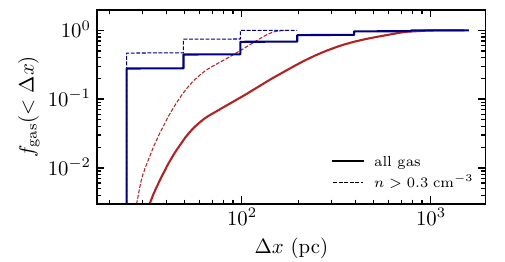}
\caption{\label{fig:resolution} Comparison of the resolution between our ART and TNG simulations at $z=3$. {\bf Top:} The relation between cell size and gas density for all cells within 30 kpc from the galaxy center. Thick, colored lines show the median values. The relation in TNG closely follows its targeted mass resolution of $8.5\times10^4\Msun$, shown with a dashed line. In ART, the scatter is significantly larger, with the median mass close to $10^4\Msun$, shown with a dotted line. {\bf Bottom:} Cumulative gas fraction as a function of cell size, including all gas (solid lines) and only dense gas with $n>0.3\cc$ (dashed lines). Although the average mass resolution is only a factor of 8 higher than in TNG (corresponding to a factor of 2 higher spatial resolution), significantly more gas mass is resolved at a higher resolution than in TNG.}
\end{figure}

Figure~\ref{fig:resolution} compares the resolution of our ART and TNG simulations using the snapshots at $z=3$. The two simulations adopt different hydrodynamic frameworks---adaptive mesh refinement in ART and moving mesh in TNG (see Section~\ref{sec:methods})---making the direct comparison challenging. 

The top panel in the figure shows the distribution of cell sizes as a function of gas density. 
In TNG, cell sizes continuously and linearly decrease with density with only a small scatter, owing to the adopted strategy of merging and splitting mesh cells to keep the gas mass within a factor of two from the target baryon resolution of $8.5\times10^4\Msun$. 
In ART, as the mesh cells are split into the octs of higher-level cells, the cell sizes change by factors of 2. The gas mass at a given level varies by orders of magnitude, with the median value roughly following a value of $10^4\Msun$, i.e., $\sim$8 times higher mass resolution, or $\sim$2 times higher spatial resolution, than in TNG.

Even though the median mass and spatial resolution as a function of density is only moderately higher in ART, significantly more gas mass is resolved down to tens of parsec scales. This is shown in the bottom panel of the figure, which compares the cumulative gas mass fraction as a function of cell size. While in TNG, the median cell size in dense gas ($n > 0.3\cc$) is $\sim$100 pc, in ART, this value is closer to 25 pc.
Note that this is mainly due to the difference in the ISM modeling: the effective EoS adopted in TNG produces a significantly higher pressure than in ART, preventing the formation of dense structures (see Section~\ref{sec:results:dense-gas}).

\subsection{ART Convergence Test}
\label{app:resolution:convergence}

\begin{deluxetable}{lccc}
\tablecaption{
The parameters of ART resimulations at different resolutions. The columns show the run label, the highest refinement level ($L_{\rm max}$), the physical cell size at $z=3$ at $L_{\rm max}$, and the gas mass threshold above which the cells at $L < L_{\rm max}$ are refined. 
\label{tab:resolution}}
\tablewidth{\columnwidth}
\tablehead{
\colhead{\hspace{.5cm}\emph{Label}\hspace{.5cm}} &
\colhead{\hspace{.5cm}\emph{$L_{\rm max}$}\hspace{.5cm}} &
\colhead{\hspace{.5cm}\emph{$\Delta x_{\rm min}$}\hspace{.5cm}} &
\colhead{\hspace{.5cm}\emph{$m_{\rm gas,max}$\tablenotemark{\scriptsize a}}\hspace{.5cm}} \\
\colhead{} &
\colhead{} &
\colhead{(pc)} &
\colhead{$(\rm M_\odot)$}
}
\startdata
fiducial & 12 & 25 & $5.70\times10^4$ \\
lowres1 & 11 & 50 & $4.56\times10^5$ \\
lowres2 & 10 & 100 & $3.65\times10^6$ \\
lowres3 & 9 & 200 & $2.92\times10^7$ \\
\enddata
\tablenotetext{\scriptsize a}{Note that the actual mass contained in cells at levels $L < L_{\rm max}$ is better than the adopted threshold, as the latter corresponds to the maximal gas mass contained in cells before the refinement condition is triggered (see, e.g.,  Figure~\ref{fig:resolution} for the median cell gas mass as a function of density in the fiducial run).}
\end{deluxetable}

\begin{figure}
\centering
\includegraphics[width=\columnwidth]{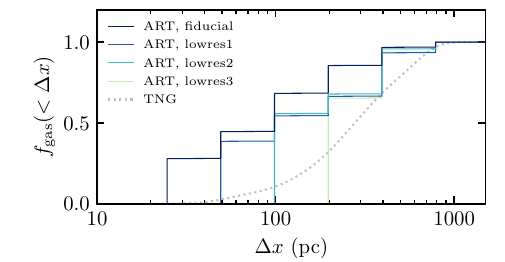}
\caption{\label{fig:art-res} Comparison of gas mass fractions contained in cells at different resolution levels in our ART convergence runs listed in Table~\ref{tab:resolution}. The dotted line shows the corresponding distribution in the TNG run for reference. This comparison is done at $z \approx 3$ using all cells within 30 kpc from the galaxy center, analogous to Figure~\ref{fig:resolution}.}
\end{figure}

\begin{figure}
\centering
\includegraphics[width=\columnwidth]{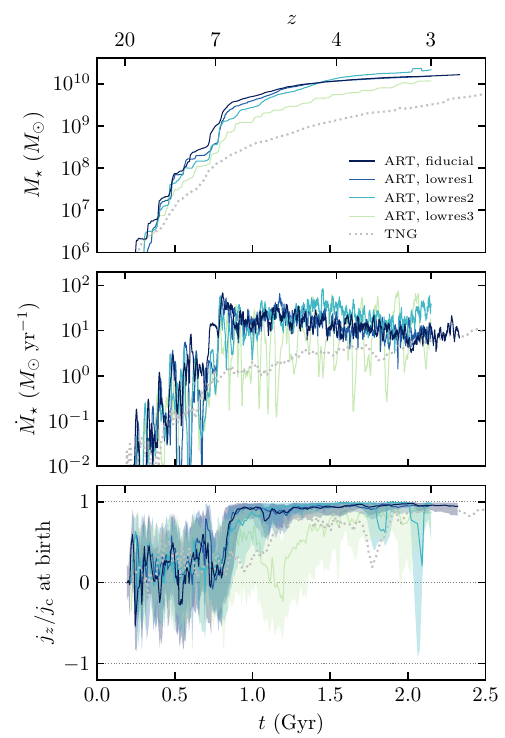}
\caption{\label{fig:art-res-sfh} Comparison of our ART convergence runs (see Table~\ref{tab:resolution}) in the analog of Figure~\ref{fig:sfh-jzjc}, which presents our main result of the paper. Adopting progressively higher resolution leads to the behavior closer to our fiducial (highest-resolution) run. The peak of the SFR (middle panel) and sudden disk formation around $z\sim 6$ (bottom panel) are clearly visible in all runs except the lowest-resolution version, ``lowres3''. }
\end{figure}

\newtext{To test the dependence of our results on resolution, we rerun our ART simulation with progressively lower resolution, reducing the spatial resolution of the highest refinement level by a factor of 2 and adjusting the gas mass refinement criterion. All runs start from a uniform $128^3$ root grid, and we progressively reduce the number of refinement levels in each subsequent run, $L_{\rm max}$, while increasing the threshold gas mass used for refinement, $m_{\rm gas,max}$, by a factor of 8. The minimal star particle mass is kept the same in all these runs, $10^4\Msun$. The resulting series of 4 runs is summarized in Table~\ref{tab:resolution}.

To quantify the difference in resolution between the resulting runs, Figure~\ref{fig:art-res} shows the cumulative gas mass fraction contained in cells at different resolution levels. For reference, the plot also shows this distribution in the TNG run. As the figure shows, the fraction of gas mass on progressively coarser levels varies by $\sim 20\%$ between the different runs, which is mainly due to the variation of the critical gas mass used to trigger refinement and the nonlinear evolution down to $z \approx 3$, when this comparison is made.

Figure~\ref{fig:art-res-sfh} is analogous to Figure~\ref{fig:sfh-jzjc} and it shows the dependence of our key results on resolution: efficient and bursty early star formation (top and middle panels) and the early formation of a star-forming disk (bottom panel). The total $\Mstar$, SFR, and the circularity of young stars remain similar when we reduce the spatial resolution by a factor of 2 (``lowres1'' run). In particular, the circularity exhibits almost identical average evolution and scatter, indicating rapid disk formation at $z\sim 6$. Further decreasing the resolution by another factor of 2 (``lowres2'' run) also leads to a qualitatively similar evolution, although with stronger deviations from our fiducial and ``lowres1'' runs. Only in our lowest-resolution run, ``lowres3'', with 8 times lower spatial resolution than that in the fiducial run, the evolution becomes substantially different, with $\jzjc$ exhibiting much larger scatter without clear settling of a thin star-forming disk. 

These results demonstrate that our main conclusions are robust for the chosen resolution. Interestingly, the behavior of global SFR and disk formation timing start converging at relatively moderate resolution: minimal cell size of $100\pc$ in the ``lowres2'' run. This can be a consequence of the explicit treatment of unresolved turbulence, which helps achieve a better accuracy and weaker sensitivity to resolution. It will be interesting to explore the role of unresolved turbulence in more detail in a separate study.}

\section{Metallicity conversion}
\label{app:FeH}

\begin{figure}
\centering
\includegraphics[width=\columnwidth]{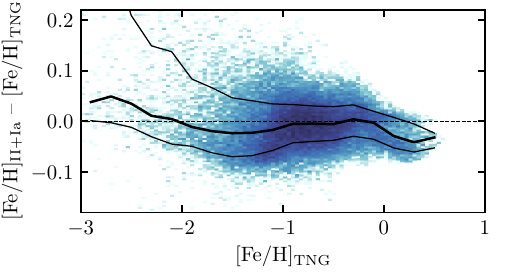}\\
\includegraphics[width=\columnwidth]{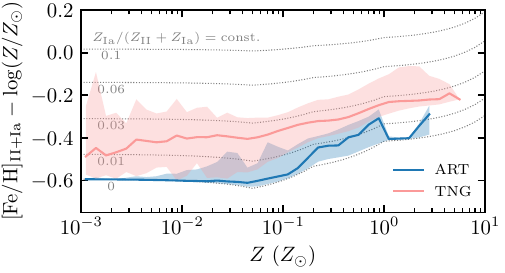}
\caption{\label{fig:FeH} Relation between stellar [Fe/H] and total metallicity, derived from the metal fractions produced by SNe Type II and Ia, $\Zii$ and $\Zia$, followed in both codes separately. The results are shown at $z=3$. {\bf Top:} The relation between the explicitly followed [Fe/H] with that derived from $\Zii$ and $\Zia$. The two agree within 0.1 dex at [Fe/H]$>-2$. {\bf Bottom:} Derived [Fe/H] as a function of total metallicity, $Z=\Zii+\Zia$. To remove the linear trend, we subtract $\log (Z/Z_\odot)$ from [Fe/H]. The dotted lines show the relations for constant SNe Ia metal fractions calculated from Equation~(\ref{eq:FeH}). The values of [Fe/H] at these early times are $\sim$0.4--0.6 dex below $Z/Z\odot$ due to the low contribution of metals injected by SNe Ia.}
\end{figure}

Observational measurements of stellar metallicities usually rely on iron abundances, [Fe/H] $\equiv (N_{\rm Fe}/N_{\rm H})/(N_{\rm Fe,\odot}/N_{\rm H,\odot})$. Here, we describe how we derive [Fe/H] from the metal mass fractions injected by SNe Type II and Ia for our analysis (e.g., in Figure~\ref{fig:FeH-pdf-spinup}). We calculate [Fe/H] from $\Zii$ and $\Zia$ because our ART situations do not explicitly track individual species, while both ART and TNG track the SNe Type II and Ia metal fractions. 

The iron abundance is calculated as

\begin{equation}
\label{eq:FeH}
    \text{[Fe/H]} = \log \left( \frac{Z_{\rm Fe}}{1-Y-Z} \frac{m_{\rm H}}{m_{\rm Fe}} \right) + 4.5 \, ,
\end{equation}
where $Z=\Zii+\Zia$ is the total metal mass fraction injected by SNe, $Z_{\rm Fe}$ is the iron mass fraction that depends on $\Zii$ and $\Zia$, $Y=0.24$ is the helium mass fraction, $m_{\rm H} = 1.01$ and $m_{\rm Fe} = 55.85$ are hydrogen and iron atomic weights, respectively, and the constant additive factor corresponds to the solar iron abundance, $\log ({N_{\rm Fe,\odot}/N_{\rm H,\odot}} ) = -4.5$. To calculate $Z_{\rm Fe}$, we use metallicity-dependent yield tables from TNG, described in \citet{pillepich18} and compute 

\begin{equation}
    Z_{\rm Fe} = \Zii\;f_{\rm Fe,II}(Z) + \Zia\;f_{\rm Fe,Ia} \, ,
\end{equation}
where $f_{\rm Fe,Ia}$ is the iron-to-metals mass fraction injected by SNe Type Ia, and $f_{\rm Fe,II}$ is the IMF-averaged iron-to-metals mass fraction of SNe Type II obtained by convolving mass- and metallicity-dependent iron and total metal yields with the \citet{chabrier03} IMF, $\xi(m) \equiv dn_\star / dm$:

\begin{equation}
\label{eq:fsnII}
    f_{\rm Fe,II}(Z) = \frac{\int m_{\rm Fe}(m,Z) \xi(m) \,dm}{\int m_{\rm Z}(m,Z) \xi(m) \,dm} \, ,
\end{equation}
with the integrals taken between the minimal and maximal SNII progenitor masses adopted in TNG, 8--120$\Msun$. The resulting values of $f_{\rm Fe,II}$ and $f_{\rm Fe,Ia}$ are listed in Table~\ref{tab:Fe}; the dependence of $f_{\rm Fe,II}$ on $Z$ is linearly interpolated between the points sampled in the table.

The resulting derived iron abundances---labeled here as [Fe/H]$_{\rm II+Ia}$---are shown in Figure~\ref{fig:FeH}. The top panel compares these derived values with the explicitly tracked iron abundances in the TNG model. The values agree within $\sim$0.1 dex. The bottom panel shows the relation between [Fe/H] and the total metallicity. At these early times, $z = 3$, [Fe/H] are $\sim$0.4--0.6 dex below $Z/Z\odot$ due to the low contribution of metals injected by SNe Ia, which produce most of the iron. Specifically, the median fractions of metal masses injected by SNe Ia by $z=3$ are $\sim 8.6 \times 10^{-3}$ in ART and $\sim 2.3 \times 10^{-2}$ in TNG. These differences reflect the differences in the star formation histories, specifically, significantly higher early SFRs in ART. 

\begin{deluxetable}{lcc}
\tablecaption{IMF-averaged iron mass fractions computed using Equation~(\ref{eq:fsnII}). Note that iron yields are significantly higher in SNe Ia, and the variation in SNe II yields across metallicity is below 40\%.
\label{tab:Fe}}
\tablewidth{\columnwidth}
\tablehead{
\colhead{} &
\colhead{\emph{SNe II}} &
\colhead{\emph{SNe Ia}}
}
\startdata
$Z = m_Z / m_{\rm \star}$ & \{0, 0.001, 0.004, 0.02\} & --- \\
$f_{\rm Fe} = m_{\rm Fe} / m_Z$ & \{0.0170, 0.0154, 0.0195, 0.0250\} & 0.541 \\
\enddata
\end{deluxetable}

\bibliographystyle{aasjournal}
\bibliography{}

\end{document}